\documentclass[pra,twocolumn,superscriptaddress,floatfix,showpacs,aps,letterpaper,nobalancelastpage]{revtex4-2}

\usepackage{graphicx} 
\usepackage{color} 
\usepackage{transparent}
\usepackage{amsmath}
\usepackage{hyperref} 
\usepackage{amssymb}
\usepackage{amsthm}
\usepackage{natbib}

\newcommand{\mycomment}[1]{}
\newcommand{\peekskill}{${\it ibm\_{}peekskill}$}

\begin{document}

\title{Characterizing non-Markovian Off-Resonant Errors in Quantum Gates}

\date{\today}

\author{Ken Xuan Wei}
\email{xkwei@ibm.com}
\author{Emily Pritchett}
\email{emily.pritchett@ibm.com}
\author{David M. Zajac}
\author{David C. McKay}
\author{Seth Merkel}
\email{seth.merkel@ibm.com}
\affiliation{IBM Quantum, IBM T.J. Watson Research Center, Yorktown Heights, NY 10598, USA}

\begin{abstract}
As quantum gates improve, it becomes increasingly difficult to characterize the remaining errors.  Here we describe a class of coherent non-Markovian errors  -- excitations due to an off-resonant drive -- that occur naturally in quantum devices that use time-dependent fields to generate gate operations. We show how these errors are mischaracterized using standard Quantum Computer Verification and Validation (QCVV) techniques that rely on Markovianity and are therefore often overlooked or assumed to be incoherent. We first demonstrate off-resonant errors within a simple toy model of Z-gates created by the AC Stark effect, then show how off-resonant errors manifest in all gates driven on a fixed-frequency transmon architecture, a prominent example being incidental cross-resonance interaction driven during single-qubit gates. Furthermore, the same methodology can access the errors caused by two-level systems (TLS), showing evidence of coherent, off-resonant interactions with subsystems that are not intentional qubits. While we explore these results and their impact on gate error for fixed-frequency devices, we note that off-resonant excitations potentially limit any architectures that use frequency selectivity.
\end{abstract}

\maketitle

\section{Introduction}
Quantum processing technologies have matured substantially enabling experiments on full sized logical qubits~\cite{Acharya2023,Sundaresan2023,zhao:2020,krinner:2020,gambetta:2022} and bringing large distance codes with small errors within reach. However, current capabilities fall short of true fault-tolerance, even if approaching that threshold. In reality, the practicalities of overhead and scale necessitate quantum gate errors lower than currently realized, thereby requiring analysis of these errors in previously untested limits. In general, improving performance over generations of quantum devices requires a real-time feedback loop between characterization, design, control, and fabrication in order to determine which (potentially very small) errors dominate systems and remove them. Within this loop, the host of errors present in realistic experiments sort conveniently into two broad categories: coherent (over-rotation, detuning, etc.) and incoherent (amplitude damping, dephasing, etc). If an error preserves coherence, it may be fixable in the control layer by clever decoupling or pulse engineering \cite{ViolaDD,KHANEJA2005,drag2009,Biercuk_2021}. Outside of generic techniques such as shortening gate times and improving circuit compilations (which are almost always already optimized with respect to the rest of a complex control trade-space), correcting incoherent errors lies in the realm of error correction or low-level hardware redesign to mitigate root cause, such as dielectric loss \cite{wang:2015,gambetta:2017}.

\begin{figure*}[ht]
\centering
\includegraphics[scale=.55]{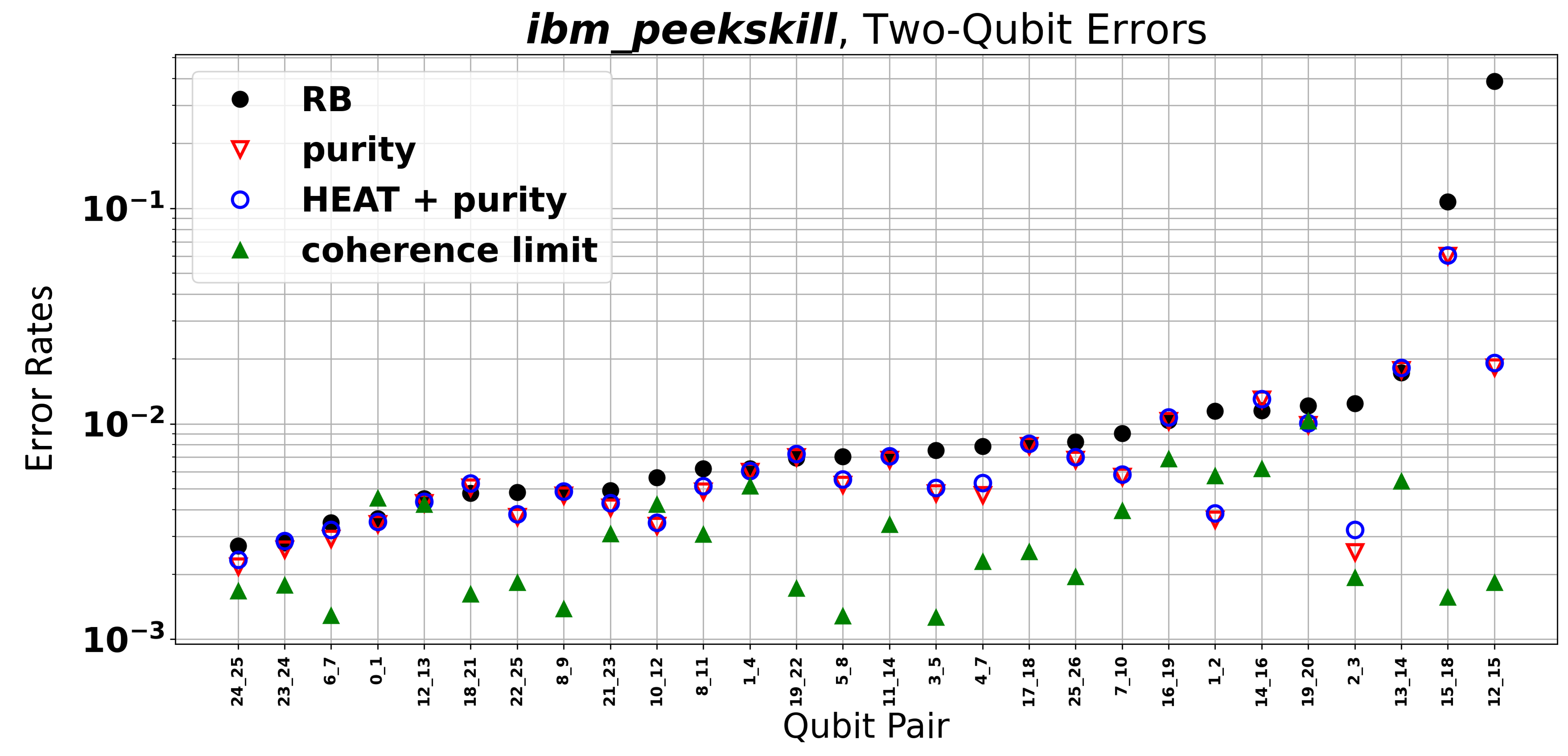}
\caption{A comparison of two qubit error rates across 27-qubit device $ibm\_peekskill$. Randomized benchmarking (RB) estimates the total error rate per two-qubit gate (black circles) sorted by error rate. The qubit pair for the measured error is listed on the x-axis where the control qubit is listed first. The gate errors are compared to the gate error from the coherence limit given by Eq.~(\ref{eqn:coh2q}) (green triangles), the error rate from purity RB (red triangles), i.e., the incoherent gate error contribution, and finally the error rates estimated by adding the purity error to the coherence error estimated by HEAT sequences (Appendix \ref{HEAT}) (blue circles).}
\label{purity}
\end{figure*}

In general, the total gate error, which we would like to partition into coherent and incoherent contributions, is measured using techniques such as randomized benchmarking (RB) \cite{Magesan12}. In RB, a random sequence of Clifford gates is applied then inverted at the end of the sequence. When averaged over many sequences, the gate error is simply related to the exponential decay of the the ground state probability as a function of the sequence length~\footnote{For two-qubit RB this procedure gives the error of the average two-qubit Clifford gate which is composed of one-qubit and two-qubit gates. Here we quote the two-qubit gate error which are the two-qubit Clifford gate error divided by 1.5, the average number of two-qubit gates per Clifford. Errors from single-qubit gates are assumed to be negligible by comparison.}. Several methodologies can be used to infer coherent/incoherent contributions to that total gate error. A zeroth-order estimate of the incoherent error  -- the ``coherence limit" (Eq.~(\ref{eqn:coh2q}) in Appendix~\ref{sect:coh}) -- can be calculated from the gate length and measured noise rates of each qubit, in particular, amplitude damping ($T_1$) and dephasing ($T_2$). Since this does not include any dynamic reduction of coherence, the coherence limit typically underestimates the error. A more robust procedure ``purity RB'' measures the purity of the state after an RB sequence \cite{Wallman_2015,mckay2016}; any difference between the RB and purity RB error rates can be ascribed to coherent errors, as discussed in Appendix \ref{sect:purity}. A number of QCVV techniques have been devised to measure coherent errors via amplification, such as Gate Set Tomography (GST) \cite{merkel:2013,Nielsen2021}, Hamiltonian Tomography \cite{sheldon2016} and Hamiltonian Error Amplifying Tomography (HEAT) sequences \cite{HEATprx2020}. The HEAT sequences we use are tailored to identify small errors to the cross-resonance interaction (see, e.g., Ref~\cite{magesan:2020}), which is the entangling mechanism utilized by the quantum processors studied in this manuscript. Originally used to amplify only those block-diagonal errors correctable with standard control parameters, here we expand HEAT to include all 15 two-qubit Pauli errors (Appendix \ref{HEAT}). \\

We use the aforementioned RB technique to measure a typical set of 2Q gate errors (measured individually) on a large quantum device -- here the 27 qubit \peekskill~device -- is presented in FIG.~\ref{purity}. 
While the errors we observe are not well correlated with the coherence limit, they track more closely with the purity error; however, discrepancies remain. Importantly, those discrepancies are not well accounted for from coherent errors that are measured from the HEAT sequences, i.e., they are not time-independent Pauli errors. Understanding possible causes of these discrepancies is the topic of this manuscript, and specifically, we focus on off-resonant errors. We will show that these errors are both coherent and invisible to our HEAT calibration techniques leading, at least in part, to the discrepancy in FIG.~\ref{purity}.\\

Off-resonant errors are both ubiquitous across many platforms for quantum information and problematic for standard, amplification-based characterization techniques like HEAT and long-sequence GST.  These errors result from frequency selectivity, a common control technique where pulses are driven at a frequency resonant with one of the many transitions of the un-driven Hamiltonian. However, due to always-on coupling in the Hamiltonian, these pulses additionally drive a number of transitions off-resonantly. While a pulse of amplitude $\Omega$ detuned from an unwanted transition (e.g. a higher transmon level or a spectator) by $\Delta$ will be suppressed to $\sim (\Omega/ \Delta)^2$, there is still residual excitation which is finite, coherent, and off-resonant with the drive pulse ~\cite{moein2022}.  Ideally $\Omega / \Delta$ is engineered so that off-resonance error rates fall below the rate of known incoherent processes (e.g., amplitude damping or dephasing) in our devices.  Unlike incoherent processes, coherent errors should be amplifiable, allowing us to verify that off-resonant errors are indeed as small as we desire and our system models are complete and accurate.  However, as we will show, off-resonant errors are invisible to HEAT, and while not strictly invisible to GST, they produce high amounts of model violation, i.e., GST completely fails to fit an error model (see Appendix \ref{GST}).  The model failure of GST hints at the underlying issue detecting off-resonant errors:  most methods in the QCVV toolbox rely on a common set of assumptions we shorthand as $Markovianity$. Concisely iterated in Ref.~\cite{knill2008}, these assumptions ensure that a gate is described by a single quantum process of fixed dimension, independent of how the gate is embedded in a larger quantum circuit. Markovianity is often invoked because it simplifies large-scale simulations and allows for rigorous mathematical statements, particularly in error correction. It is similarly a core assumption for many experimental QCVV protocols, even though most physical processes exhibit some level of non-Markvoian behavior. When Markovianity is assumed erroneously, coherent, non-Markovian errors can be mis-characterized as incoherent, Markovian errors; this mistake prevents us from pinpointing error mechanisms and engineering solutions. For example, non-Markovianity caused by drifts in device parameters, which may arise due to a number of sources such as magnetic field drifts in neutral atom and ion traps, and drifts of control electronics in most quantum technologies, require different control solutions \cite{Proctor2020,Biercuk_2021} than $T_1$ decay due to dielectric loss.\\ 

Our paper is organized as follows. In \S~\ref{sect:rotframes}, we describe how off-resonant errors break the stationary assumption of Markovianity which states that gate errors must be independent of when the gate is performed. We show that for off-resonant errors this assumption depends on the choice of rotating frame, leading to the situation where for any frame only some, but not all, of the gates in a gate set may have Markovian descriptions. We then propose a practical alternative to GST or HEAT for amplifying and detecting these errors. By carefully interleaving frame changes into amplification sequences, off-resonant errors can be made to add constructively. Next, in \S~\ref{Sec:Stark}, we investigate this concept experimentally with a well-controlled version of an off-resonant error which occurs when using a Stark tone to make a $Z$ gate. Unless otherwise stated, the experiments in this paper were performed on an internal IBM device with specifications similar to ${\it ibm\_peekskill}$.  In \S~\ref{sect:cnot} we explore off-resonant errors in two-qubit gates that are generated from the cross-resonance interaction. These gates will necessarily have some off-resonant error because the control qubit is driven off-resonantly at the target qubit frequency~\cite{moein2022}. We show how Derivative Removal by Adiabatic Gate (DRAG) compensation pulses~\cite{chow2011,moein2022} can be used to mitigate these errors and achieve an error rate of 1.3e-3, comparable to the lowest measured in similar CR systems~\cite{maple2021,sizzle2022}.  Finally, in \S~\ref{sect:spect} we explore how spectator qubits are driven off-resonantly during single qubits gates due to always on coupling. We also observe that these spectator ``qubits" are sometimes described by spurious TLS as opposed to the engineered systems in our processors.  While a small effect,  spectator errors will eventually become bottlenecks as other error sources are improved, revealing the rich physics in our devices.

\section{Off-resonant errors and continuous phase amplification \label{sect:rotframes}}
In this section we show that the stationary assumption of Markovianity is frame-dependent, and we introduce an amplification-based characterization technique sensitive to non-stationary errors.  To show how non-Markovian errors manifest we start with a toy model of the Hamiltonian of a single qubit  with two drives,
\begin{align}
H(t) &= -\frac{\omega_{\rm q}}{2} Z\nonumber\\ 
&+ \Omega_0(t) \cos([\omega_{\rm q}+\Delta_0] t) X\nonumber\\
&+\Omega_1(t) \cos([\omega_{\rm q}+\Delta_1] t) X, \label{ModelHam}
\end{align}
with qubit frequency $\omega_{\rm q}$, drive detunings $\Delta_{0,1}$, and envelope functions  $\Omega_{0,1}(t)$.  Even with a single drive, $\Omega_1 = 0$, this Hamiltonian exhibits non-stationary behavior.  Imagine a family of envelope functions displaced in time by $T_0$, i.e., $\Omega_0(t;0) = \Omega_0(t+T_0,T_0)$, The resulting Hamiltonian, 
\begin{align}
&H(t+T_0;T_0) \nonumber\\
&= -\frac{\omega_{\rm q}}{2} Z + \Omega_0(t+T_0,T_0) \cos([\omega_{\rm q}+\Delta_0] (t+T_0)) X\nonumber\\
&= -\frac{\omega_{\rm q}}{2} Z + \Omega_0(t,0) \cos([\omega_{\rm q}+\Delta_0] (t+T_0)) X\nonumber\\
&\neq H(t;0), \label{ModelShiftedHam}\end{align}
is not stationary because the carrier does not share the envelope's symmetry.  When we move to the frame rotating at $\omega_{\rm q}+\Delta_0$ and take the rotating wave approximation (RWA) dropping terms at $\pm 2 (\omega_q + \Delta_0)$, $H(t) \rightarrow \frac{\Delta_0}{2} Z + \frac{\Omega_0(t)}{2} X$, and the stationary property is restored, i.e., it is only counter rotating terms that are non-stationary in this frame.  If counter-rotating terms were the only source of non-stationary processes in our gates it would be safe to make the stationary assumption; however it is more complicated when we have multiple drives at different frequencies.  While possible to find a rotating frame where any individual term is stationary, it may be impossible to find a single rotating frame where the entire Hamiltonian exhibits stationary behavior.  Even if $\Omega_0$ and $\Omega_1$ describe non-overlapping pulses starting at $T_0$ and $T_1$ respectively, the resulting unitary evolutions integrated over the non-zero domains of the envelopes, $U_0[T_0]$ and $U_1[T_1]$, cannot both be Markovian unless the two drive frequencies are commensurate.  If we choose a frame where $U_0[T_0]$ is independent of $T_0$, i.e., $U_0[T_0]=U_0$ is stationary, then in that frame $U_1[T_1] = U^\dagger_{\rm rot}[T_1+t_{\rm g}] U_1 U_{\rm rot}[T_1]$, where $U_1=U_1[T_1]$ in its stationary frame, $U_{\rm rot}[T] =e^{-i \frac{(\Delta_1-\Delta_0) T}{2} Z}$ transforms the frame where $U_0$ is stationary to the frame where $U_1$ is stationary at time $T$, and $t_{\rm g}$ is the duration of $U_1$.  Clearly this operator depends on $T_1$ and therefore is non-stationary unless $U_1$ commutes with $U_{\rm rot}$, and while sometimes ideal gates commute with frame transformations, their errors may not.  Furthermore, Markovianity does not appear to be a property of a given gate, since either gate can be expressed as Markovian in the proper frame, but is instead a holistic property of the gate set.

We can now express why non-stationary, off-resonant errors can be tricky to quantify using calibration routines for high-fidelity gates (e.g. HEAT) that amplify coherent errors to fine-tune any free parameters in (\ref{ModelHam}).  In the stationary case, repeated coherent errors grow quadratically, and we can design SPAM-free fitting routines. A typical amplification experiment goes as follows:  prepare a superposition of eigenstates of the un-driven Hamiltonian, repeat application of a gate a number of times $N$, then measure some observable in the energy eigenbasis.  

However, to amplify errors that anti-commute with $U$, it is neccesary to interleave some other interrogation gate, $V$,  constructing sequences of the form $(UV)^N$.  While $U$ might have a stationary description, $V$ might not be stationary in the same frame, and in general, \emph{there may be no single choice of rotating frame in which all relevant gates are Markovian simultaneously}. Consider amplifying errors in the gate $U_1$ from above. We know there is a stationary frame for $U_1$, so we can amplify those gate errors that commute with $U_1$ by repeating it $n$ times in sequence. Since the frame is arbitrary, we can consider this amplification sequence in any rotating frame,
\begin{align}
   &U_1[n t_{\rm g}] \ldots U_1[t_{\rm g}]U_1[0]\nonumber\\
   & =  U^{\dagger}_{\rm rot}[(n+1)t_{\rm g}] U_1 U_{\rm rot}[n t_{\rm g}]\ldots U^{\dagger}_{\rm rot}[2t_{\rm g}] U_1 U_{\rm rot}[t_{\rm g}] U^{\dagger}_{\rm rot}[t_{\rm g}] U_1  \nonumber \\
    & =  U^{\dagger}_{\rm rot}[(n+1)t_{\rm g}] U_1^n,
\end{align}
and note that there is still coherent amplification. However, if we try and amplify a mixed set of $U_0$ and $U_1$ gates, where for example we express the sequence in $U_0$'s stationary frame and define all gates to have duration $t_{\rm g}$, we get
\begin{align}
    &U_0[(2 n-1) t_{\rm g}]U_1[(2n-2) t_{\rm g}]\ldots U_0[t_{\rm g}] U_1[0] \nonumber \\
    & =  U_0 U^{\dagger}_{\rm rot}[(2 n-1) t_{\rm g}] U_1 U_{\rm rot}[(2n-2)t_{\rm g}] \ldots U_0 U^{\dagger}_{\rm rot}[t_{\rm g}] U_1  \nonumber \\
    & =  \prod_{j=0}^{n-1} U_0U^{\dagger}_{\rm rot}[(2j+1)t_{\rm g}] U_1 U_{\rm rot}[2j t_{\rm g}]. \label{eqn:mixed}
\end{align}
Assuming that $U_{\rm rot}$ does not have a period commensurate with $t_{\rm g}$, not only does this sequence not amplify errors in $U_1$, it instead suppresses them in an average sense as the phase is essentially randomized from one application to the next.  \\

Fortunately, techniques like HEAT and GST can be recovered for a scenario like Eq.~(\ref{eqn:mixed}) by absorbing the rotating frame into an interleaved gates (i.e., by interleaving a non-stationary gate); however this can be tedious. A simpler approach is to use an interleaved gate that commutes with the rotating frame.  Projected onto a qubit, commuting interrogation gates are simply $Z$-rotations which can be implemented a variety of ways such as tuning the qubit energy, interleaving a delay, decomposing a $Z$-rotation into the standard single-qubit gates, or by performing $Z$-gates in software by updating the phase of subsequent operations \cite{mckay2017}.  Phase updates $\phi_{\rm d} \rightarrow \phi_{\rm d} + \phi$ can be expressed as the transformation $U \rightarrow  e^{-i  \phi H_{\rm phase}} U e^{i  \phi H_{\rm phase}}$ for some appropriate choice of $H_{\rm phase}$. Incrementing the phase $\phi$ between successive applications of $U$ results in the unitary 

\begin{align}
   U_{\rm amp} &=
    \left(e^{-i (n-1) \phi H_{\rm phase}} U e^{i (n-1) \phi H_{\rm phase}}  \right)\ldots\nonumber\\
    &\times \left(e^{-i  \phi H_{\rm phase}} U e^{i  \phi H_{\rm phase}}  \right)   U, \label{2Dsweep}
\end{align}
which can be reordered to make apparent the typical form of an amplification experiment interleaved with a diagonal interrogating gate: 
\begin{equation}
U_{\rm amp}=e^{-i n \phi H_{\rm phase}} \left(e^{i \phi H_{\rm phase}} U  \right)^n.
\label{cpa}
\end{equation}
We call sweeping the phase $\phi$ in Eq.~(\ref{cpa}) \emph{continuous phase amplfication}, a technique we find broadly useful for identifying the off-resonant errors that can limit our device's performance. We will explore the use of this technique in the following sections. \\

\section{Stark Z-gates}\label{Sec:Stark}

\begin{figure}[thb!]
\includegraphics[width=.4\textwidth]{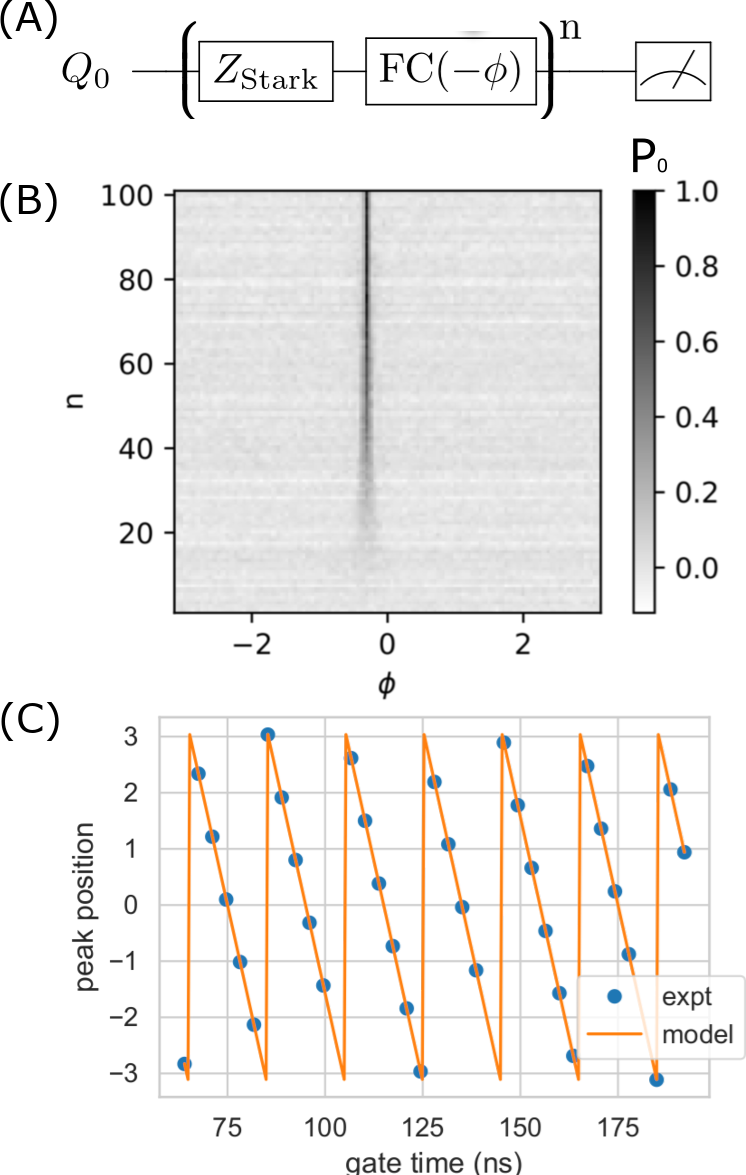}
\caption{({\bf A}) Continuous phase amplification of Stark $Z_{\pi/2}$ gate errors with an interrogation frame change (FC) of phase increment, $\phi$. {(\bf B)}. shows the excitation probability of $Q_0$ (labeled as $P_0$) as a function of the phase increment and the number of repetitions with a total population inversion around $n=100$.  ({\bf C}) The measured phase of maximal error amplification (expt) has good agreement with the model predictions from Eq.~(\ref{eq:starkpeak}).}
\label{Stark1D_OC}
\end{figure}

\begin{figure*}[thb!]
\includegraphics[width=0.8\textwidth]{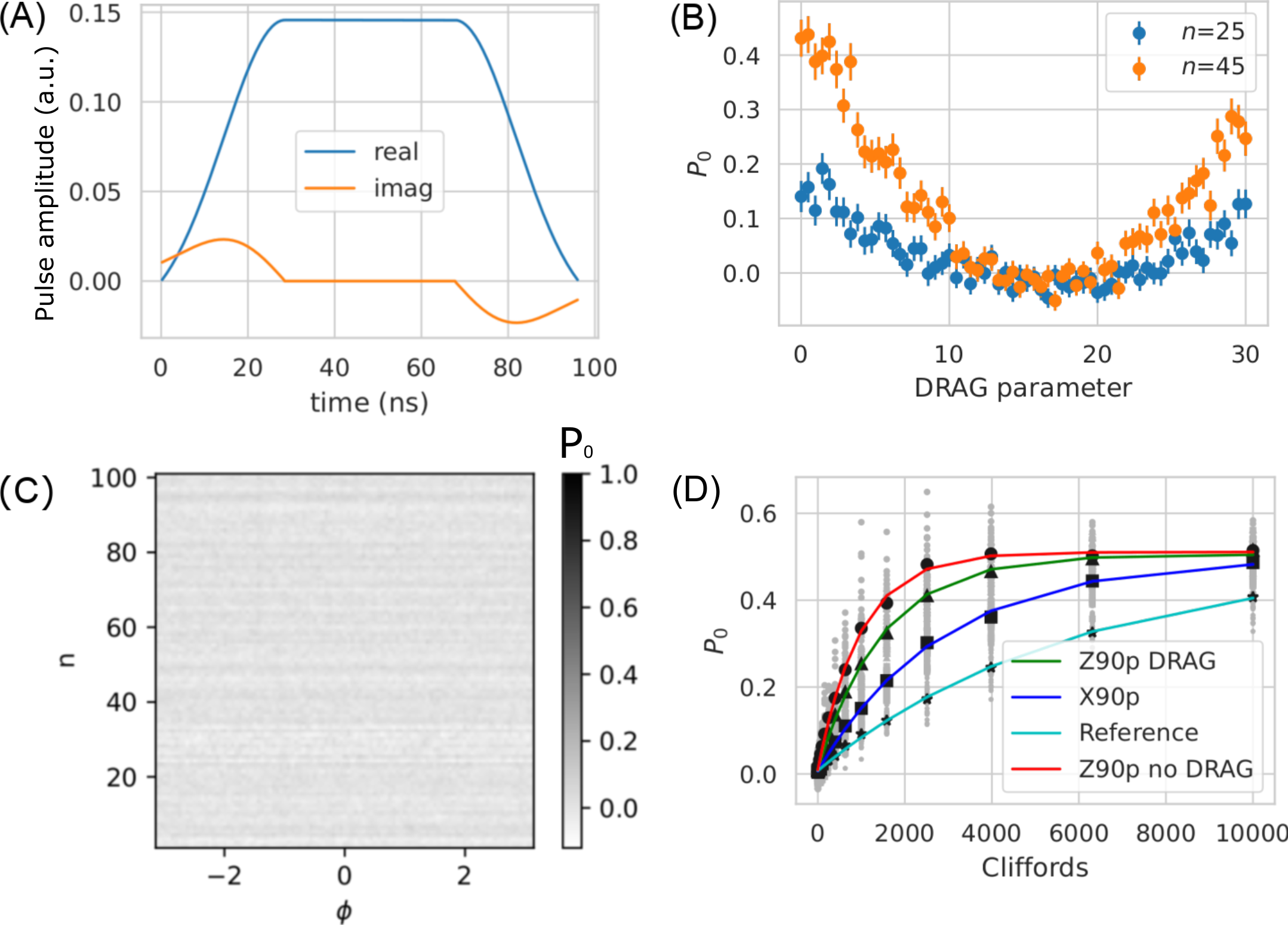}
\caption{({\bf A}) Drag correction to a Gaussian square pulse.  ({\bf B}) The continuous phase amplification sequence, at $\phi_{\rm peak}$, shows a clear minima for the excitation probability $P_0$ for both $n=25$ and $n=40$ at the optimal DRAG amplitude.  ({\bf C}) For these DRAG parameters the residual off-resonant excitation from FIG.~\ref{Stark1D_OC} (B) vanishes in continuous phase amplification.  ({\bf D}) The resulting error of the Stark $Z_{\pi/2}$ gate as measured by interleaved randomized benchmarking drops from 4.36e-4 to 2.52e-4 with DRAG (details in main text).}
\label{Stark1D_DRAG_OC}
\end{figure*}

Now we consider perhaps the simplest example of off-resonant error one can generate on a driven qubit (such as the transmons considered here): the off-diagonal corrections to a $Z_{\pi/2}$ gate  generated by driving a two-level system off-resonantly, that is, a diagonal gate implemented by a Stark shift. The model Hamiltonian (\ref{ModelHam}) becomes
\begin{equation}
    H(t)= -\frac{\omega_{\rm q}}{2}Z  +\Omega(t) \cos(\omega_{\rm d} t-\phi_{\rm d}) X
\end{equation}
when projected onto a single qubit. To make $H$ stationary we choose rotating frame  $H_{\rm rf} = \frac{\omega_{\rm d}} {2}Z$, which, after the RWA, leads to the effective Hamiltonian
\begin{equation}
    H'(t)= \frac{\Delta}{2}{Z}+\frac{\Omega(t)}{2}\big( \cos{\phi_d} X + \sin{\phi_d} Y \big)\label{Eq:Stark}
\end{equation}
with detuning $\Delta \equiv \omega_{\rm d}-\omega_{\rm q}$.
In the limit where $\Omega(t) \ll \Delta$, $H'$ generates a rotation that is only slightly perturbed from the Z-axis.  To lowest order and in the square pulse approximation, this perturbation changes the effective Z-rotation  from $\theta_Z = \Delta t_{\rm g}\rightarrow\Delta t_{\rm g} + \frac{\Omega^2 t_{\rm g}}{2 \Delta}$ over gate duration $t_{\rm g}$; therefore in a frame rotating with the qubit energy, the resonant frame, we see an effective $Z$-rotation of $\theta_{\rm Stark} =  -\frac{\Omega^2 t_{\rm g}}{2 \Delta}$.  In addition to the $Z$-rotation due to Stark shift, there are non-Markovian errors since the rotation axis for the off-resonant drive is slightly tilted from Z. Rotation around the tilted Z-axis does not commute with rotation around original Z. In the qubit's resonant frame the tilted Z-axis is also rotating, giving rise to time-dependent errors that are difficult to detect in experiments such as Rabi oscillations.
In order to measure them directly with Hamiltonian tomography (i.e. process tomography as a function of $t_g$) we need to resolve the excitation rate
\begin{equation}
P_{10}\equiv \vert \langle 1 \vert U_{\rm eff}(t_g) \vert 0 \rangle \vert^2 = \frac{\Omega^2}{\Omega_{\rm r}^2} \sin^2\left(\frac{ \Omega_{\rm r} t_g }{2}\right)\label{eq:P10} 
\end{equation}
where $U_{\rm eff}(t_g)=e^{-i (\Omega X + \Delta Z) t_g/2}$ and $\Omega_{\rm r}^2=\Omega^2 + \Delta^2$.
The maximum contrast for this signal is $\sim \left(\Omega/\Delta\right)^2$, which has no $t_{\rm g}$ dependence, requiring that our resolution/shots scale with the expected bare error rate.  Repetition doesn't amplify these excitation errors since they anti-commute with the dominant $Z$-rotation.  We could try to amplify with interrogating $X$ or $Y$ gates, but these resonantly-driven gates are stationary in the \emph{resonant} frame where off-resonant excitations are non-stationary, destructively interfering on average. In section IV, we will show that the off-resonant error in Stark Z-gates is an important source of coherent error in CNOT gates using cross-resonance drives. High fidelity physical Z-gates, such as the Stark Z-gates described in this section, will be important in quantum circuits where virtual Z-gates cannot be moved across two-qubit interactions such as the $\sqrt{\text{ISWAP}}$ gate.  \\ 

We have described a simple off-resonant error that is both hard to amplify using standard methods of characterization/calibration and treated as negligible under the Markovian assumption -- a dangerous combination especially as error rates approach ``the fault tolerant threshold" requiring unprecedented levels of precision.  Fortunately, as is the case for any coherent error, there must be $\emph{some}$ choice of amplification sequence that results in quadratic growth of off-resonant errors. Our solution is to use the continuous phase amplification technique described in the previous section. We demonstrate this experimentally with the Stark $Z_{\pi/2}$ gate generated by a $t_g=96$ ns Gaussian square pulse detuned $\Delta = -50$ MHz below the qubit's resonance (qubit parameters given in TABLE~\ref{starkparams}).  This test case provides a very clean, accessible demonstration of off-resonance physics at gate times and detunings that are comparable to the off-resonant driving of cross-resonance gates discussed in the next sections.\\

Continuous phase amplification of the Stark $Z_{\pi/2}$ is performed by the pulse sequence shown in  FIG.~\ref{Stark1D_OC} (\textbf{A}). We perform a two-dimensional sweep over the total number of repetitions, $N$, and the phase incremented in between repetitions, $\phi$, as the interrogating frame change (FC).   The measured excited state probability inverts completely in less than 100 repetitions, as shown in FIG.~\ref{Stark1D_OC} (\textbf{B}). This non-negligible error would be invisible to standard characterization techniques for which $\phi=0$.  The peak location is centered around $\phi_{\rm peak}$ for which  $e^{i \phi_{\rm peak} H_{\rm phase}} U$ from (\ref{2Dsweep}) is purely equatorial, that is, the phase update -- which acts like a $Z$-rotation in the qubit subspace -- completely cancels the $Z$-component of the gate leaving only the small $X$ error.  Repetition then amplifies this error giving a large signal (albeit for only a narrow range of $\phi$ values). Zeroing the $Z$-component of  $e^{i \phi_\text{peak} H_{\rm phase}} U$, i.e.,  setting  $\text{Tr}(Ze^{i \phi_\text{peak} H_{\rm phase}} U)\propto\frac{\Delta}{\Omega_r}\cos(\phi_\text{peak}/2)\sin(\Omega_r t_\text{g}/2)-\cos(\Omega_r t_\text{g}/2)\sin(\phi_\text{peak}/2)=0$, then expanding with respect to the small parameter $\Omega /\Delta$ gives 
\begin{align} \label{eq:starkpeak}
    \phi_\text{peak}\simeq\text{sign}(\Delta)\Omega_r t_g= \Delta t_g +\theta_\text{Stark},
\end{align}
where we have used the definition of Stark shift $\text{sign}(\Delta)\omega_\text{Stark} = \sqrt{\Omega^2+\Delta^2} - |\Delta|$ and $\theta_\text{Stark}=\omega_\text{Stark} t_g$. As shown in FIG.~\ref{Stark1D_OC} (\textbf{C}), we see excellent agreement between predicted and experimentally extracted peak positions for different $t_{\rm g}$ demonstrating the robustness of Eq.~(\ref{eq:starkpeak}). Note that for a few gate times we did not observe a peak as $\text{mod}(\Omega_r t_g,2\pi)\simeq 0$ and excitation errors vanish.  Our understanding of this particular error makes a 2D sweep to discover $\phi_{\rm peak}$ unnecessary; however, in more complex examples, it's not practical to predict the phase of one (or more) peaks $a$ $priori$.\\

\begin{table}[t]
\begin{ruledtabular}
	\begin{tabular}{|c c|}
	$f_{01} \ ({\rm GHz})$ &  $5.165$   \\
	$\alpha$ ({\rm MHz}) &   $-346$   \\
	$f_\text{\rm readout} \ ({\rm GHz}) $ & $7.083$ \\
	$T_1$ \ ($\mu$s) &   $124(6)$  \\
	$T_{2\rm echo}$\ ($\mu$s) &   $107(8)$  \\
	$Z_{\pi/2}$ EPG &   $4.36$e-$4\pm1.8$e-$5$   \\
	$Z_{\pi/2 \rm DRAG}$ EPG &   $2.52$e-$4\pm9.3$e-$6$   \\ 
	\end{tabular}
\end{ruledtabular}
\caption{Summary of parameters describing the Stark $Z_{\pi/2}$ gate.}
\label{starkparams}
\end{table}

Now that we have a tool to measure off-resonant errors we can design control sequences to correct them.  A small $Y$-rotation will correct this off-resonant error which has rotated a $Z_{\pi/2}$ gate only slightly in the $X$-$Z$ plane.  Crucially, this $Y$-rotation has to be in phase with the Stark gate.   Instead of phase-matching a positive Y-pulse before the gate and a negative Y-pulse after, we take advantage of DRAG \cite{drag2009,chow2011}, FIG.\ref{Stark1D_DRAG_OC} (\textbf{A}) -- by definition a derivative pulse that is in-phase with the Stark pulse up to a $\pi/2$ offset.  DRAG doesn't lengthen the gate's duration and only has one free parameter to calibrate, its relative amplitude. The amplified excitation at $\phi_{\rm peak}$ shows a clear minima with respect to DRAG amplitude for different numbers of repetition, as shown in FIG.~\ref{Stark1D_DRAG_OC} (\textbf{B}). Continuous phase amplification of the Stark gate including an optimized DRAG pulse shows a dramatic reduction of off-resonant errors (FIG.~\ref{Stark1D_DRAG_OC} (\textbf{C})), nearly halving the $Z_{\pi/2}$ error as measured by interleaved randomized benchmarking (FIG.~\ref{Stark1D_DRAG_OC} (\textbf{D})). The gate set used in the reference RB is comprised of: $X_{\pm \pi/2}$, $Y_{\pm \pi/2}$, $Z_{\pm \pi/2}$, and $Z_{0,\pi}$; where all $Z$ gates are implemented via software frame changes \cite{mckay2017} and $X/Y$ gates are Gaussian pulses with gate time $4\sigma$ and $\sigma\approx7.11$ns. The interleaved gates are $X_{\pi/2}$, 96ns Stark $Z_{\pi/2}$ with DRAG correction, and 96ns Stark $Z_{\pi/2}$ without DRAG correction. The Stark Z gates use flat-topped Gaussian pulses where rise and fall are $2\sigma$ long with $\sigma\approx 14.22$ns.

\section{Non block-diagonal CNOT errors \label{sect:cnot}}

\begin{figure*}[thb!]
\centering
\includegraphics[width=0.8\textwidth]{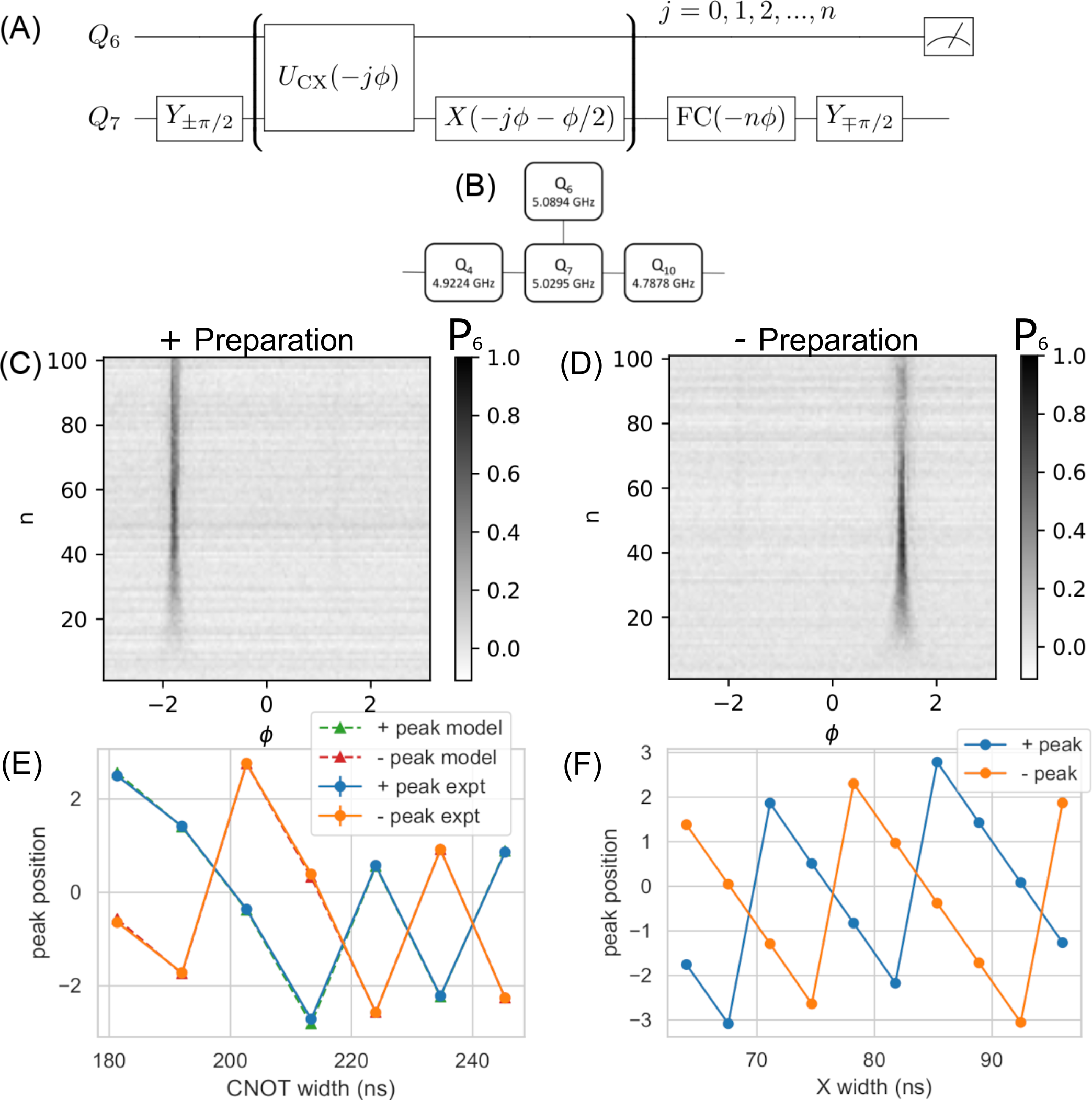}
\caption{State-selective frame spectroscopy amplifies CR errors using an interrogating frame change, tracking the eigenstates of the frame-shifted CR pulse with a rotated $X$-pulse {\bf (A)}.  We drive CR tones on the control qubit $Q_6$ resonant with the target qubit $Q_7$ ({\bf B}). For the two target eigenstates $\vert +\rangle$ and $\vert -\rangle$ we observe peaks in the excitation probability ($P_6$) of the control qubit, as shown in ({\bf C}) and ({\bf D}).  The observed peak positions (expt) fit well to the expression in Eq.~(\ref{eq:cnotpeaks}) (model) where there is a dependence on both the length of the CNOT ({\bf E}) and the length of the $X_\pi$ pulse following the CNOT ({\bf F}). }
\label{CNOT_OC}
\end{figure*}

\begin{figure*}[thb!]
\centering
\includegraphics[width=0.8\textwidth]{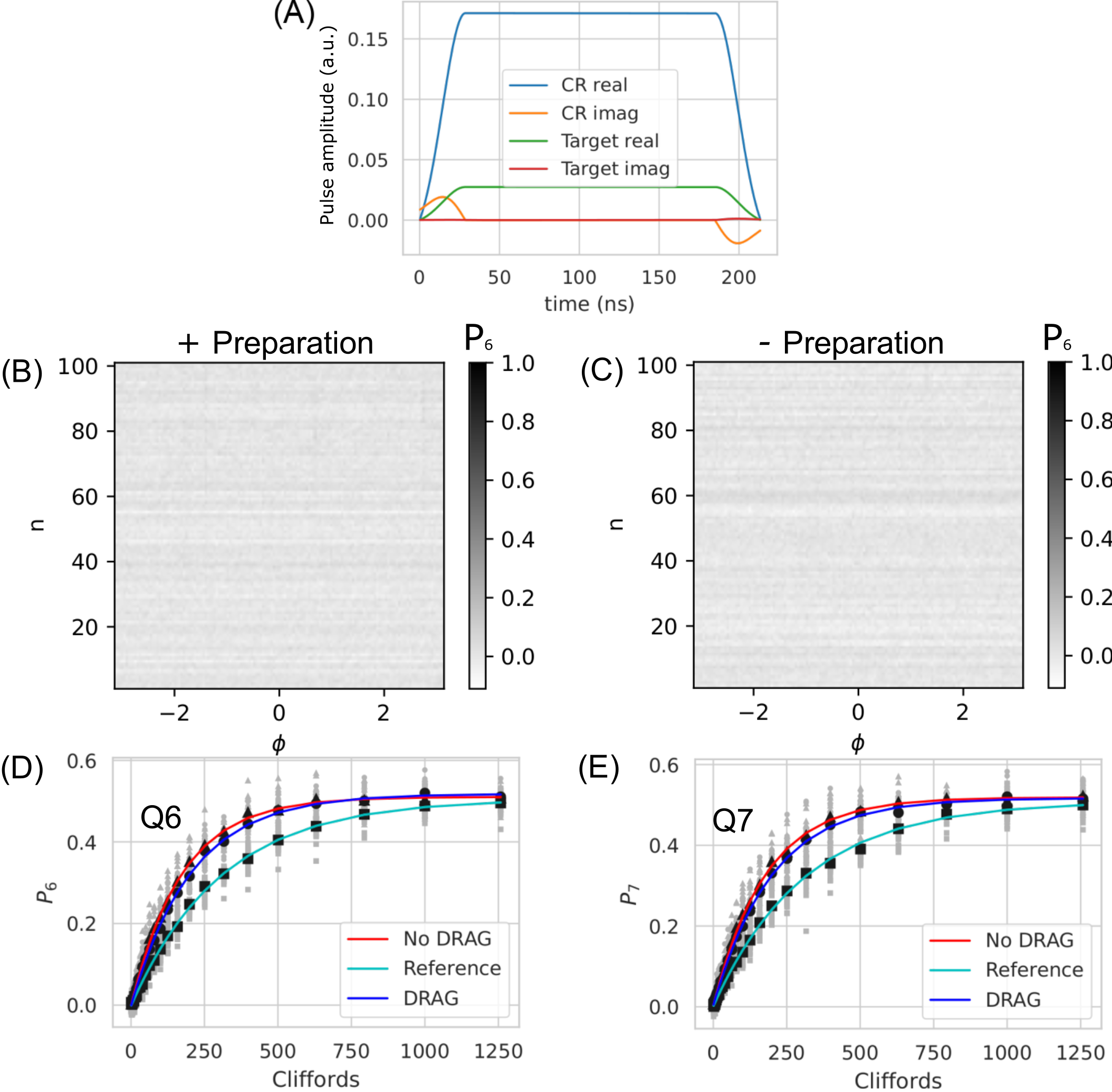}
\caption{({\bf A}) Optimized control pulse including DRAG and target rotary tone.  In ({\bf B}) and ({\bf C}) we see that once again the off-resonant excitation in the control qubit ($Q_6$) is suppressed for both target eigenstates $\vert +\rangle$ and $\vert -\rangle$.  The DRAG correction reduces the error as measured by interleaved RB from 1.8e-3 to 1.2e-3 (({\bf D}) and ({\bf E}), details in main text); $P_6$ and $P_7$ are excitation probabilities of $Q_6$ and $Q_7$ respectively.}
\label{CNOT_DRAG_OC}
\end{figure*}

Given the results of the previous section, can we apply them to improve two-qubit gates? Fortunately, the Stark example of the previous section relates directly to the CNOT gate implemented by cross-resonance (CR) where the control qubit is driven far off-resonance at the target qubit's dressed frequency.  The control qubit undergoes a Stark shift due to this drive  \cite{magesan:2020}, and while the resulting $Z$-rotation is corrected either by echo or with single-qubit gates, additional off-resonant excitation errors are present and potentially undetected.\\

Expressed in the qubits' dressed basis, the full CR Hamiltonian is given by $H_{\rm CR} = H_0 + \Omega(t) \cos(\omega_{\rm target} t + \phi) H_{\rm drive}$, where to leading order $H_0=\omega_{\rm control} ZI/2 + \omega_{\rm target} IZ/2 + \zeta ZZ$ and $H_{\rm drive} = XI + \mu ZX + \nu IX$.  Here $\mu$ and $\nu$ describe the cross-resonance and cross talk-terms respectively \cite{magesan:2020,moein2020}. While $ZZ$-errors are a topic of ongoing research at IBM \cite{sheldon2016,sizzle2022}, we ignore the $ZZ$-term in the following arguments as it is fairly small and  straightforward to detect by conventional techniques.  Because the drive Hamiltonian contains both an $XI$-term  and a $ZX$-term, the only choice of frame that preserves Markovianity rotates both qubits at the drive (target) frequency:  $H_{\rm RF} = \omega_{\rm target} \left(ZI + IZ \right)/2$.
After performing the RWA, 
\begin{align}
H_{\rm CR}(t)& = -\frac{\Delta}{2} ZI \nonumber\\
&+ \frac{\Omega(t)}{2} \cos(\phi_d) \left(XI + \mu ZX +\nu IX \right) \nonumber\\
&+  \frac{\Omega(t)}{2} \sin(\phi_d) \left(YI + \mu ZY +\nu IY \right)
\end{align}
where  $\Delta\equiv\omega_{\rm target}-\omega_{\rm control}$.  We must transform the control qubit to its resonant frame in order for the resonantly-driven single-qubit gates to be stationary. While this frame change commutes with the CNOT gate itself, it will cause all errors that aren't block-diagonal with respect to the control qubit to become non-stationary (and thus hard to detect with HEAT).\\

To construct a CNOT from the CR interaction we typically set phase $\phi_d=0$, choose an envelope that integrates a $ZX_{\pi/2}$, and correct the $ZI$ and $IX$ coefficients with single qubit gates and/or active cancellation.  Gate calibration routines evolve, as thoroughly described in \cite{maple2021,sizzle2022,HEATprx2020,QV64_2022}; however, usually we neglect the off-resonant $XI$ contribution (similarly to the naive treatment of the Stark gate) due to its small magnitude at desirable detunings.  Naively executing a continuous phase amplification experiment as described in the previous section generates a strong signal of excitation from the $XI$ off-resonant term, but it can be more difficult to interpret.  Consider a unitary gate derived from a square pulse CR gate,   
\begin{align}\label{eq:crham}
    U_{CR}&=e^{-i H_{CR} t_{\rm g}}=U_+ \otimes|+\rangle\langle+| + U_- \otimes|-\rangle\langle-| \nonumber\\
    &=e^{-i \left[\Omega X - (\Delta-\mu\Omega)Z +\Omega\nu I \right]t_{\rm g}/2}\otimes|+\rangle\langle+|\nonumber\\
    &+ e^{-i \left[\Omega X - (\Delta+\mu\Omega)Z - \Omega\nu I \right]t_{\rm g}/2}\otimes|-\rangle\langle-|
\end{align}
where $t_{\rm g}$ is now the CNOT gate time, and $U_\pm$ are the control unitaries when the target qubit is in the $|\pm\rangle$ states.  The interrogating frame change rotates both the control and the target qubits -- which are defined to be in the same frame for CR --  rotating off-resonant errors on the control qubit as desired, but also scrambling $U_+$ and $U_-$ and leading to complicated dynamics. In order to probe  off-resonant errors in $U_+$ and $U_-$ individually we need to undo the phase on the target qubit due to the interrogating frame update. To do so we prepare the target in  $|+\rangle/|-\rangle$ states, which are eigenstates parallel/anti-parallel to the very first CR pulse of the amplification sequence. Then, before each repetition of the CR pulse with incremented phase $\phi$, we perform an $X_\pi$-pulse with its phase incremented by $\phi/2$ from the previous CR-pulse, rotating the eigenstates to those of the next CR pulse.  This amplification sequence, shown in FIG.~\ref{CNOT_OC} (\textbf{A}), keeps the target qubit state parallel or anti-parallel with the rotated CNOT gate, allowing the control qubit to independently evolve according to $U_+$ or $U_-$ and effectively decoupling the two sectors of off-resonant errors.  Adapting Eq.~(\ref{eq:starkpeak}) for the CR gate we expect peak positions
\begin{align}\label{eq:cnotpeaks}
    \phi_{\pm\text{peak}} = \text{sign}(\Delta \mp \mu \Omega)\Omega_{\mp r} t_g = \Delta t_g - \theta_{\pm \text{Stark}} 
\end{align}
where $\Omega_{\pm r}=\sqrt{\Omega^2 + (\Delta\pm\mu\Omega)^2}$ are the Rabi rates and $\theta_{\pm \text{Stark}}$ are the rotations due to Stark shifts when the target is in the $|\pm\rangle$ states, respectively. Experimentally we observe one peak in the control qubit when the target is in the $|+\rangle$ state and another peak, separated by $\pi$, when the target is in the $|-\rangle$ state, as shown in FIG.~\ref{CNOT_OC} (\textbf{C}) and (\textbf{D}). This $\pi$-separation is a signature of CNOT. We refer to this particular application of continuous phase amplification as \emph{state-selective frame spectroscopy.} \\
\begin{table}
    \begin{ruledtabular} 
        \begin{tabular}{|ccc|} 
            Parameters & $Q_6$ & $Q_7$\\
            \hline
            $T_1$ ($\mu$s) & 333(25) & 324(22) \\
            $T_{2\text{echo}}$ ($\mu$s) & 313(45) & 271(29) \\
            $f_{01}$ (GHz) & 5.089 & 5.030 \\
            $f_\text{\rm readout} \ ({\rm GHz})$ & 7.394 & 7.142 \\
            $\alpha$ (MHz) & -343 & -343 \\
            CNOT EPG & $0.00191 \pm 9.9$e$-5$ & $0.00176 \pm 9.8$e$-5$ \\
            $\text{CNOT}_\text{DRAG}$ EPG & $0.00125 \pm 8.3$e$-5$ & $0.00124\pm 9.3$e$-5$
        \end{tabular} 
    \end{ruledtabular}
    \caption{Parameters describing the two qubits used in the off-resonant CNOT experiment as well as resulting error rates.}
\label{paramcnot}
\end{table}

One subtlety with this type of experiment is that the $t_{\rm g}$ variable in Eq.~(\ref{eq:cnotpeaks}) should really denote the repetition time of the CNOT plus the interrogating $X$-pulse. In FIG.~\ref{CNOT_OC} (\textbf{F}), the peak positions extracted at a fixed number of repetitions $n=45$ vary linearly with  $X$-gate time with slope $\Delta$, as expected.  We evaluate $\sin(\phi_{\pm\text{peaks}})$ and fit to $\sin(a t_{X} + \phi_0)$, taking $\phi_0$ is the `extrapolated' peak position if the $X$-pulse had no duration. In FIG.~\ref{CNOT_OC} (\textbf{E}) we compare the extrapolated peak positions with Eq.~(\ref{eq:cnotpeaks}) for different CNOT gate times and observe good agreement. The rotations due to Stark shifts $\theta_{\pm\text{Stark}}$ are measured using Ramsey experiments on the control qubit when the target is prepared in $|\pm\rangle$ state.  Like in the previous section, we can optimize the DRAG parameter to suppress the off-resonant excitation in the control qubit for the CNOT gate~\cite{moein2022}. As shown in FIG.~\ref{CNOT_DRAG_OC} (\textbf{B}) and (\textbf{C}), an optimized DRAG pulse eliminates both off-resonant peaks apparent in state-selective frame spectroscopy. The optimized pulse shape is shown in FIG.~\ref{CNOT_DRAG_OC} (\textbf{A}), and we observe an improvement in two-qubit error obtained from interleaved randomized benchmarking for the DRAG optimized CNOT, as shown in FIG.~\ref{CNOT_DRAG_OC} (\textbf{D}) and (\textbf{E}). The estimated error per gate (EPG) for CNOT with DRAG correction is 1.2e-3, and for CNOT without DRAG is 1.8e-3. We extract the EPGs from exponential fits to the averages of 18 different RB sequences, shown by the lines in FIG.~\ref{CNOT_DRAG_OC} (\textbf{D}) and (\textbf{E}). The CNOT gates use 213.33ns long flat-topped Gaussian pulse where rise and fall are $2\sigma$ with $\sigma\approx 14.22$ns. We use the CNOT calibration procedure described in \cite{maple2021,sizzle2022}, where different pulse parameters are simultaneously calibrated. The reference RB sequence uses the following gate set: $X_{\pm \pi/2}$, $Y_{\pm \pi/2}$, $Z_{\pm \pi/2}$, $Z_{0,\pi}$, and DRAG corrected CNOT. The $Z$ gates are virtual frame changes, and $X/Y$ gates are Gaussian pulses $4\sigma$ long with $\sigma\approx 7.11$ns. TABLE~\ref{paramcnot} summarizes the qubit parameters and gate errors, the detuning for this gate is $|\Delta| \approx 60$ MHz.

\section{Single qubit spectator errors \label{sect:spect}}
\begin{figure*}[thb!]
\centering
\includegraphics[width=1.0\textwidth]{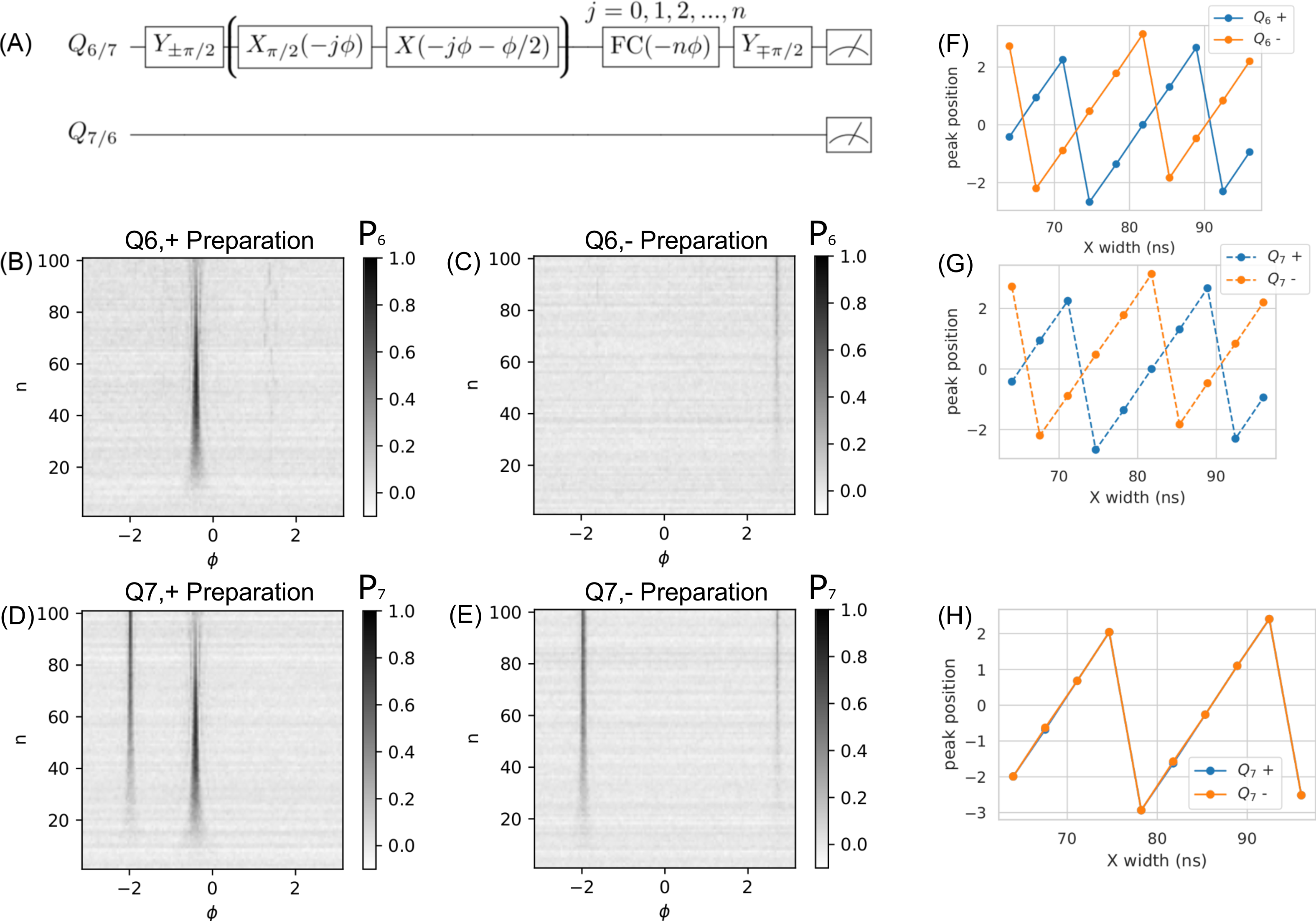}
\caption{State-selective frame spectroscopy of an $X_{\pi/2}$ gate ({\bf A}) on either qubit $Q_6$ or $Q_7$ with the other qubit treated as a spectator.  In ({\bf B}) and ({\bf D}), the excitation probabilities for the driven qubit $Q_6$ ($P_6$) and for spectator qubit $Q_7$ ($P_7$) are measured when $Q_6$ is in put in the $\vert + \rangle$ state. In ({\bf C}) and ({\bf E}), the excitation probabilities are measured when $Q_6$ is in put in the $\vert - \rangle$ state. In ({\bf B}) and ({\bf D}), there is a common peak in $Q_6$ and $Q_7$ appearing at the same location which is separated from the common peak in ({\bf C}) and ({\bf E}) by $\pi$, as expected from Eq.~(\ref{eqn:specpeak}). The lone peak in ({\bf D}) and ({\bf E}) is the $IX$ peak. The dependence of the common peak position on the length of the $X_\pi$ pulse are shown in ({\bf G}) and ({\bf F}) respectively for $Q_6$ and $Q_7$. The dependence of the $IX$ peak on $X_\pi$ length is shown in ({\bf H}).}
\label{spec_1}
\end{figure*}

\begin{table*}
    \begin{ruledtabular} 
        \begin{tabular}{|ccccc|}
            Qubits & $\phi_{\pm\text{peak}}(t^*_X=0)$ & $\Delta t_{\rm g} \pm \frac{\pi}{2}$ & $\phi^{IX}_{\pm\text{peak}}(t^*_X=0)$ & $\Delta t_{\rm g}$\\
            \hline
            $Q_6$ (Driven qubit) & $0.662,-2.480$ & $0.640,-2.502$ &  & \\
            $Q_7$ (Spectator qubit) & $0.662,-2.480$ & $0.640,-2.502$ & $-0.969,-0.887$ & $-0.931$ 
        \end{tabular} 
    \end{ruledtabular}
    \caption{Correlated peaks $\phi_{\pm {\rm peak}}$, extrapolated to zero duration $X$-pulse $t^*_X=0$, on the driven and spectator qubit separated by a comma for the two different preparations: $|+0\big>$, $|-0\big>$.  This agrees well with the prediction from Eq.~(\ref{eqn:specpeak}) (2nd column). Similarly, a spectator qubit peak  varies only slightly with state preparation, its observed mean (extrapolated) position $\phi^{IX}_{\pm\text{peak}}(t^*_X=0)$ agrees well with $\Delta t_{\rm g}$. }
    \label{paramspec67}
\end{table*}

When neighboring qubits are subject to cross-talk, either classically through the control lines or quantum crosstalk through a fixed coupling, resonant drives can generate off-resonant errors. While these errors will show up in simultaneous RB~\cite{gambetta:2012}, we would instead like to amplify and detect them directly.  Here we use state-selective frame spectroscopy to detect off-resonant errors in spectator qubits resulting from single qubit gates.  Consider a simplified version of the full Hamiltonian from the previous section where we now drive at the frequency of the control qubit (now labeled qubit 0), and treat the target as a spectator.  When we move to the frame where both qubits oscillate at the drive frequency we obtain
\begin{align*}
    H_{\rm sq}=\frac{\Omega}{2} (XI + \mu ZX + \nu IX) -\frac{\Delta}{2} IZ
\end{align*}
where $\Delta=\omega_{\rm spectator}-\omega_{\rm q}$.  In the basis of $|+0\rangle$, $|-1\rangle$, $|-0\rangle$, and $|+1\rangle$
\begin{align}\label{eq:specham}
    H_{\rm sq}=\frac{1}{2}
    \begin{bmatrix}
    -\Delta+\Omega & \mu \Omega & 0 & \nu\Omega \\
    \mu \Omega & \Delta - \Omega & \nu\Omega & 0 \\
    0 & \nu\Omega & -\Delta-\Omega & \mu \Omega \\
    \nu\Omega & 0 & \mu \Omega & \Delta + \Omega \\
    \end{bmatrix}
\end{align} is diagonally dominant so long as $|\Delta-\Omega|>>\mu\Omega,\nu\Omega$.  
  Entangling and classical cross-talk are generated between the driven qubit and its spectator at rates determined by $\mu$ and $\nu$ respectively, and both of these errors will be off-resonant with the spectator qubit.

As an intuition building gedanken experiment, consider a single qubit $X_{\pi/2}$ gate generated by a square pulse where we further assume $\nu=0$.  Like the CR case, the Hamiltonian is block-diagonal, with $|+0\big>$
only interacting with $|-1\big>$, and $|-0\big>$ only interacting with $|+1\big>$.
The pulse sequence is shown in FIG.~\ref{spec_1} (\textbf{A}) will amplify entanglement errors due to $\mu\neq 0$ as the CR case, modified only in that the rotating $X$-pulse preserves the eigenstates of the primary driven qubit (instead of the target).  Thanks to the state-selective nature of our pulse sequence, by preparing the initial state in either $|+0\rangle$ or $|-0\rangle$ we  probe the two off-resonant errors independently. Using the same analysis as the two previous sections, we expect correlated peaks in both the driven and spectator qubits whose positions are given by
\begin{align}
    \phi_{\pm\text{peak}} = \text{sign}(\Delta \mp \Omega)\Omega_{\mp r} t_g \approx (\Delta \mp \Omega)t_g = \Delta t_g \mp \frac{\pi}{2}
\label{eqn:specpeak}
\end{align}
where the Rabi rates are $\Omega_{\pm r}=\sqrt{\mu^2\Omega^2 + (\Delta\pm\Omega)^2}\approx|\Delta \pm \Omega|$, and we have neglected terms proportional to $\mu^2$ since $\mu$ is small. Experimentally we observe peaks near these predicted value ($\phi=-0.42$) in both the driven qubit $Q_6$ and spectator qubit $Q_7$ when the initial state is in $|Q_6 Q_7\rangle = |+0\rangle$, as shown in FIG.~\ref{spec_1} (\textbf{B}) and (\textbf{D}), and $\pi$ away ($\phi=2.72$) when prepared in $|-0\rangle$, as shown in FIG.~\ref{spec_1} (\textbf{C}) and (\textbf{E}).\\ 

In addition to peaks found on both the driven and spectator qubits, another peak appears only on the spectator qubit (at $\phi=-1.99$) regardless of the initial state of the driven qubit. This results from $\nu IX\neq 0$, which leaves the Hamiltonian  non-block-diagonal. Including this term makes calculating the peak positions analytically more difficult; however, since $\mu$ and $\nu$ are relatively small, the peak positions can be estimated independently.  Since the $IX$ peak is due to an off-resonant error only on the spectator qubit, its peak position is given simply by $\Delta t_{\rm g}$. The positions of the common peaks depend linearly on $X$-gate duration with slope given by $\Delta$, as shown in FIG.~\ref{spec_1} (\textbf{F}) and (\textbf{G}), as does the position of the peak only appearing in the spectator qubit $Q_7$, as shown in FIG.~\ref{spec_1} (\textbf{H}). Using the same technique as in the CNOT section, we extrapolate the peak positions $\phi_{\pm {\rm peak}}$ in the limit of a zero-duration $X$-pulse and compare with our model Eq.~(\ref{eqn:specpeak}). The results are summarized in TABLE~\ref{paramspec67}.\\

\begin{figure*}[thb!]
\centering
\includegraphics[width=.8\textwidth]{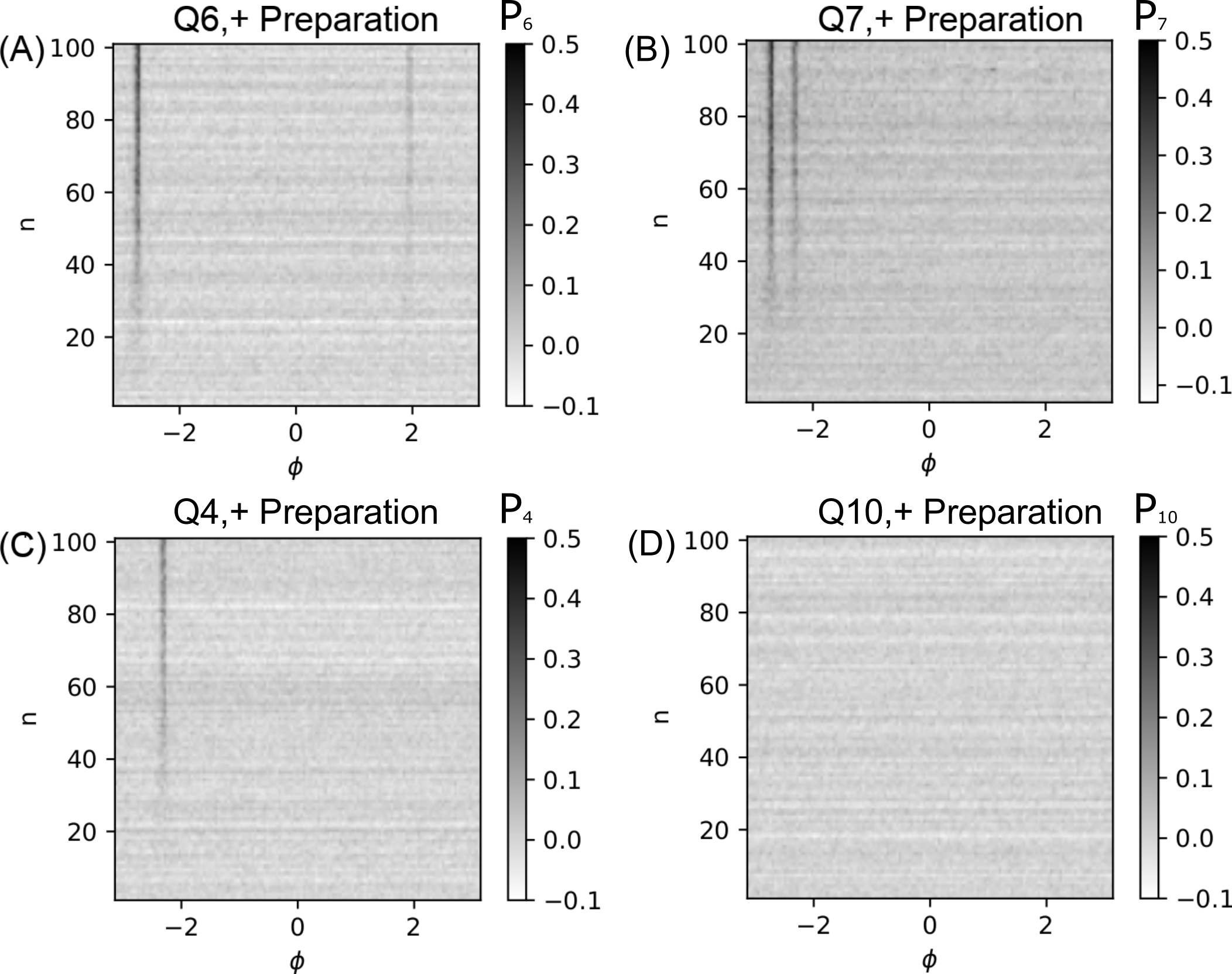}
\caption{State-selective frame spectroscopy of an $X_{\pi/2}$ gate performed on the driven qubit $Q_7$ with $Q_4$, $Q_6$, and $Q_{10}$ as spectator qubits as seen in FIG.~\ref{CNOT_OC}({\bf B}).  In all cases $Q_7$ is prepared in $\vert + \rangle$ and we measure the excitation probabilities $P_6$ ({\bf A}), $P_7$ ({\bf B}), $P_4$ ({\bf C}) and $P_{10}$ ({\bf B}). In ({\bf A}) and ({\bf B}) there is a common peak in $Q_6$ and $Q_7$ appearing at the same location, and a lone $IX$ peak in $Q_6$. In ({\bf B}) and ({\bf C}) there is a different common peak in $Q_7$ and $Q_4$, and no $IX$ peak in $Q_4$. Lastly in ({\bf D}) there are no peaks at all, since $Q_{10}$ is far detuned from the driven qubit $Q_7$.}
\label{spec_2}
\end{figure*}

\begin{table*}
    \begin{ruledtabular}
        \begin{tabular}{|ccccc|} 
            Qubits & $\phi_{+\text{peak}}(t_{X}=64\text{ns})$ & $\Delta (t_{\rm g}+t_{X}) + \frac{\pi}{2}$ & $\phi^{IX}_{+\text{peak}}(t_{X}=64\text{ns})$ & $\Delta (t_{\rm g}+t_{X})$\\
            \hline
            $Q_7$ (Driven qubit) & $-2.723,-2.304$ & $-2.732,-2.328$ & & \\
            $Q_6$ (Spectator qubit) & $-2.723$ & $-2.732$ & $1.990$ & $1.980$ \\
            $Q_4$ (Spectator qubit) & $-2.356$ & $-2.328$ & & 
        \end{tabular} 
    \end{ruledtabular}
    \caption{Correlated peaks $\phi_{\pm {\rm peak}}$ on the driven and spectator qubit separated by a comma for the two different preparations: $|+0\big>$, $|-0\big>$.  This agrees well with the prediction from Eq.~(\ref{eqn:specpeak}) with $t_{\rm g}\rightarrow t_{g}+t_{X}$ (2nd column). Similarly, a spectator qubit peak  varies only slightly with state preparation agreeing well with $\Delta(t_{\rm g}+t_{X}$).}
    \label{paramspec76}
    \end{table*}

Next we repeat the experiments driving $Q_7$ so that $Q_6$ is the spectator. For the initial state $|Q_7 Q_6\rangle = |+0\rangle$, we observe a peak (at $\phi=-2.7$) on both $Q_7$ and $Q_6$ and a $IX$ peak (at $\phi=1.99$) on $Q_6$. The 2D sweep for $Q_6$ is shown in FIG.~\ref{spec_2} (\textbf{B}). Notice that the peaks in $Q_6$ and the peaks in $Q_7$ in FIG.~\ref{spec_1} (\textbf{E}) are reflections of each other with respect to $\phi=0$. This makes sense since in our model exchanging the driven and spectator qubit amounts to changing $\Delta\rightarrow -\Delta$. However the amplitudes of the peaks are less for $|Q_7 Q_6\rangle$ than $|Q_6 Q_7\rangle$. For the driven qubit $Q_7$, in addition to the peak shared by $Q_6$, there is another peak (at $\phi=-2.2$) shared with another spectator qubit $Q_4$ shown in FIG.~\ref{spec_2} (\textbf{C}). Indeed $Q_7$ has three spectator qubits with $Q_6$ being the closest in detuning (60 MHz), followed by $Q_4$ (107 MHz) and $Q_{10}$ (241 MHz). We did not observe any spectator excitation on $Q_{10}$, as shown in FIG.~\ref{spec_2} (\textbf{D}). The qubit connectivity and frequencies are shown in FIG.~\ref{CNOT_OC} (\textbf{B}).  In addition to obtaining the peak positions based on extracting the $t_X \rightarrow 0$ limit, one can also keep $t_X$ as it is and modify Eq.~(\ref{eqn:specpeak}) as $\phi_{\pm \text{peak}}=\Delta(t_{\rm g} + t_X) \mp \frac{\pi}{2}$. We use this method to analyze the peak positions observed for the case where $Q_7$ is the driven qubits. As shown in TABLE~\ref{paramspec76}, we again see good agreement between experiments and model.\\

The $X_{\pi/2}$ pulses used in this section are Gaussian pulses with length $t_{X90}=4\sigma$ and $\sigma=3.55$ns. We used much longer pulses, $64$ns, for the $X$ gate in the sequence to minimize spectator off-resonant errors during that pulse since we are focusing on spectator errors due to the $X_{\pi/2}$ gate. Our analysis in this section indicates that one need not drive at an amplitude comparable to detuning to induce appreciable spectator errors; in fact, an excited qubit may exchange or swap excitation with a spectator of small detuning even in the absence of drive  \cite{zajac2021}.  These errors are off-resonant and therefore easy to overlook.  Here the $X_{\pi/2}$ pulses had an average drive amplitude of 17 MHz, which is less than 1/3 of the frequency difference between $Q_6$ and $Q_7$. Yet remarkably, the spectator error after 30 $X_{\pi/2}$ pulses on $Q_6$ is enough to put the driven and spectator qubits into a Bell state. We also point out that the entangling interaction is similar to the FLICFORQ gate described in Ref.~\cite{rigetti2005}. More worryingly our experiments show that single qubit gates can introduce entangling errors with multiple spectator qubits, even if they are detuned by 100 MHz or more. \\

We point out the key observation in section \ref{sect:cnot} and \ref{sect:spect} is that by preparing in specific initial states and keeping track of the phase of the interleaved $X_\pi$ pulse (see FIG. ~\ref{CNOT_OC}A and \ref{spec_1}A), the off-resonant errors in cross-resonance and spectator can be reduced to analyzing the dynamics of two independent single-qubit systems, instead of one two-qubit system. This simplification allows us to directly obtain Eq.~(\ref{eq:cnotpeaks}) and Eq.~(\ref{eqn:specpeak}) by reading off the Hamiltonian (Eq.~(\ref{eq:crham}) and Eq.~(\ref{eq:specham})) using the same reasoning leading to Eq.~(\ref{eq:starkpeak}) for the Stark Z gates. We want to emphasize that section \ref{Sec:Stark}, \ref{sect:cnot}, and \ref{sect:spect} are closely related, and we are essentially analyzing the same problem in slightly different settings. \\

Unfortunately there is no simple fix to these off-resonant spectator errors using known pulse optimizations such as DRAG.  For current devices and fidelity goals, we are able to work in regimes of large enough detuning to mostly ignore these off-resonant errors on single qubit gates.  Designing pulse sequences to correct these errors is out of the scope of this current manuscript, but could prove an important future area of research if we find we need to relax constraints on detuning or lower single qubit gate errors way below the current $10^{-4}$ levels.  For the remainder of this section we will explore two additional examples of off-resonant errors in single qubit gates. 

\subsection{Off-resonant errors and CPMG}

\begin{figure*}[thb!]
\centering
\includegraphics[width=.8\textwidth]{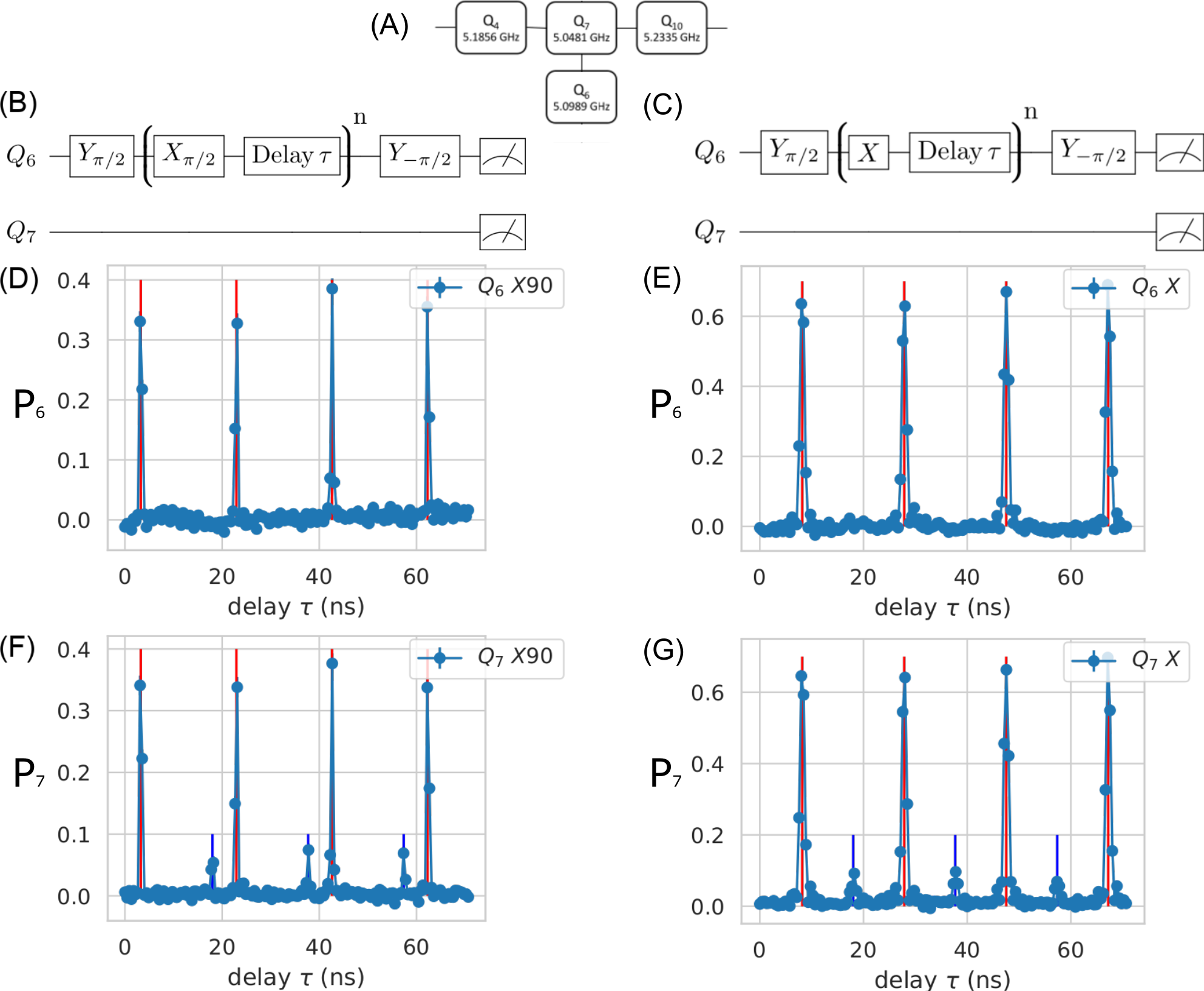}
\caption{Amplifying spectator errors with CPMG sequences for $X_{\pi/2}$ ({\bf B}) and $X_\pi$ ({\bf C}) gates on $Q_6$ in ({\bf A}).  We measure excitation probability $P_6$ in ({\bf D}, {\bf E}) and $P_7$ in ({\bf F}, {\bf G}) for a $X_{\pi/2}$ gate ({\bf D}, {\bf F}) and $X_{\pi}$ gate ({\bf E}, {\bf G}). The red vertical lines show the expected positions of the entangling peaks, and the blue vertical lines show the expected positions of spectator ($IX$) peaks. The entangling peaks appear at the same location in both qubits. These pulse sequences are also used in Quantum noise spectroscopy (QNS) and dynamical decoupling magnetometry. The key difference is that we are not using the pulse sequence to probe noise or field during the delay, instead we are using the delay to probe and amplify the spectator error in the pulse.}
\label{dd_OC}
\end{figure*}

Instead of continuous phase sweeps, it is possible to observe off-resonant errors in the more standard framework of dynamical decoupling (DD) sequences. Consider the pulse sequences shown in FIG. (\ref{dd_OC}) ({\bf B},{\bf C}), where the driven qubit is initially prepared in $|+\rangle$ state, followed by a periodic application of either an $X_{\pi/2}$ or $X_{\pi}$ interleaved by a delay $\tau$. In the case of an $X_{\pi}$-pulse this is the well-known CPMG sequence. As shown in FIG. (\ref{dd_OC}), we observe excitation in both the driven and spectator qubits at regular intervals separated by $2\pi/|\Delta|$. These peaks are actually the same as those observed before in frame spectroscopy experiments.  We can use $\phi_\text{peak}=\Delta \tau_\text{peak}$ and directly obtain the peak positions as $\tau_\text{peak}=-t_g-\theta_g + 2 m \pi/|\Delta|$, where $m$ is an integer and $\theta_g = \pi/2$ for $X_{\pi/2}$ pulses and $\theta_g=\pi$ for $X_{\pi}$ pulses. On the spectator qubit we see another set of peaks. These are the $IX$ peaks, and their positions are given by $\tau_\text{peak}=-t_g + 2 m \pi/|\Delta|$. Observing off-resonant excitation peaks in DD is limited by the sampling resolution in the delay $\tau$. With a large repetition number $n$ the peak width can be quite small, one can easily lose the peaks if the resolution in $\tau$ is not small enough. For this set of data we used a different device $\it{ibmq}\_\it{cairo}$ where the electronics allowed us to sweep $\tau$ in increments of $0.222ns$ (a 16$\times$ shorter increment than available on the other two devices used). The qubit parameters are shown in FIG.~\ref{dd_OC} ({\bf A}). Here both $X_{\pi/2}$ and $X$ gates are Gaussian pulses $4\sigma$ long with $\sigma=5.33$ns; the repetition numbers are $n=16$ for $X$ and $n=32$ for $X_{\pi/2}$. We point out that compared to $\tau$, the minimum phase increment on IBM deployed hardware is almost infinitesimal.  Resolving these errors for any but the smallest detunings is much more practical using phase sweeps. We note a recent work describing non-Markovian effects in single qubit gates in a transmon processor \cite{Li2023}.

\subsection{Evidence of TLS in continuous phase sweeps}

So far we have used continuous phase sweeps to ascertain parameters in otherwise extremely well-understood Hamiltonian models.  However, in superconducting qubits loss of coherence is typically described by coupling to other, less well-understood two-level systems (TLS) typically thought to be due to microscopic irregularities in the device.  The dynamics of TLS near the qubit can be probed via Stark spectroscopy \cite{Carroll2022}, spin-lock spectroscopy \cite{abdu_2020,leonid2023prxq}, and careful study of Ramsey sequences \cite{gulacsi2023}.  

\begin{figure*}[thb!]
\centering
\includegraphics[width=.8\textwidth]{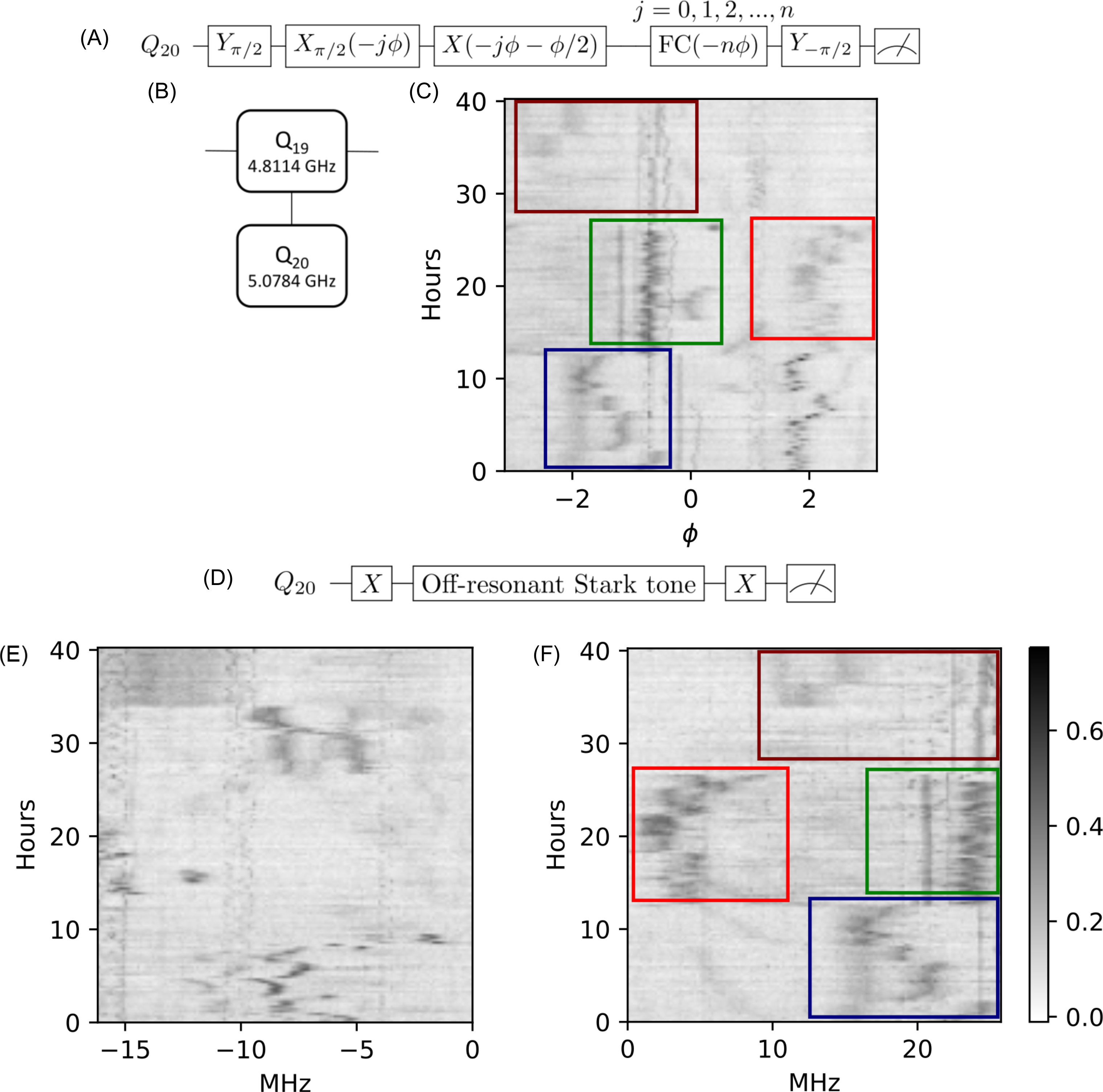}
\caption{We compare the features in a continuous phase sweep ({\bf A},{\bf C}) to those observed in Stark spectroscopy ({\bf D}) for an 80MHz drive tuned above ({\bf E}) and below ({\bf F}) the qubits resonance frequency.  This data was take on $ibm\_whiplash$ and the observed lines in the phase plot are not predicted by spectator interactions as the neighboring qubit ({\bf B}) is far detuned.  Several of the features in the phase sweep are observed in the Stark spectroscopy (highlighted by the colored boxes as a guide to the eye), and so both experiments are probing TLS physics.}
\label{tls_OC}
\end{figure*}

Here we point out that frame spectroscopy developed for observing off-resonant errors can also reveal TLS dynamics. We treat the TLS exactly as we do spectator qubits, although they are typically lower coherence.  In FIG.~\ref{tls_OC}({\bf A}) we apply the $X_{\pi/2}$ amplification sequence shown in FIG.~\ref{spec_1}({\bf A})  for a large number of repetition number ($n=1000$), then compare to Stark TLS spectroscopy described in \cite{Carroll2022} where excited state population is measured after 20 $\mu$s of off-resonant drive 80MHz above/below the qubit frequency FIG.\ref{tls_OC}({\bf F}/{\bf G}). The resulting spectra are monitored every 15 minutes for over 40 hours. The qubit used is far detuned ($\sim 260$ MHz) from its only neighbor to minimize the effects of spectator errors described in the previous section. As shown in FIG.~\ref{tls_OC}, both Stark TLS and frame spectroscopy display peaks which can move with time.  While some peaks from Stark spectroscopy can be associated with peaks in the phase plots, it is clear that the phase plot reveal a much richer structure than one gets from $T_1$ and $T_2$ measurements alone.

\section{Conclusion}

In this manuscript we demonstrated a broad category of ``off-resonant excitation errors" that can show up in common gates utilized on quantum processors and are difficult to quantify with standard QCVV techniques. The main issue is that the assumptions of Markovianity (the stationary assumption in particular) are violated when the gates in a gate set (and their errors) are stationary only in incommensurate rotating frames. This recasts Markovianity not as an intrinsic property of a single gate, but rather as a statement about an entire quantum processor.  We developed an alternative method for characterizing these errors by using interrogation gates (described as phase updates, $Z$-rotations, or frame changes) that are co-Markovian with the gate being measured.  This allows us to detect, and in many cases mitigate, off-resonant excitations on multiple gates of a standard fixed-frequency IBM device. In the cases studied here, the magnitude of these errors are just on the cusp of becoming performance bottlenecks \cite{Li2023}, which is perhaps not surprising if we consider the long-term efforts of optimizing device parameters with respect to metrics such as randomized benchmarking and quantum volume \cite{cross2019,QV64_2022}.  Without an understanding of the physical mechanism of the errors that affect large-scale metrics they can only be mitigated through the slow process of trial and error in device design and fabrication, or by decreasing drive amplitudes (increasing gate times) until their rates become comparable to those of the background incoherent processes. With new protocols, as described here, we hope that we can begin to better understand this wide variety of off-resonant errors and begin the process of engineering corrections so that they are never the limiter of performance. While DRAG worked for certain scenarios in this manuscript, the most ubiquitous -- spectator errors during single qubit gates -- remains uncorrected and an open question going forward. Also, as the field pushes on techniques to reduce errors algorithmically, such as error mitigation \cite{Kim2023,vandenBerg2023} and quantum error correction, understanding the strength of these off-resonant errors and how they impact these protocols is of the utmost importance. Our study should be of immediate interest to quantum noise spectroscopy (QNS) \cite{Bylander2011,murphy2022}, quantum signal processing \cite{low2017}, and dynamical decoupling based magnetometry \cite{liu2019}, where off-resonant errors, if not account for, could lead to spurious results. Furthermore, the experimental techniques developed here could be applied to study many-body resonances in many-body localized systems (MBL) \cite{Berke2022,morningstar2022}, potentially shedding light to the stability of MBL systems.\\   

\begin{acknowledgments}
The authors would like to thank Luke Govia, Kentaro Heya, Abhinav Kandala, Isaac Lauer, Moein Malekakhlagh, George Stehlik, Neereja Sundaresan, and Matthew Ware for insightful discussions and Will Shanks for software support. The devices used in this work were designed and fabricated internally at IBM. This work was supported by IARPA under LogiQ (contract W911NF-16-1-0114) and the Army Research Office under QCISS (W911NF-21-1-0002). All statements of fact, opinion or conclusions contained herein are those of the authors and should not be construed as representing the official views or policies of the US Government.
\end{acknowledgments}

\appendix
\section{Coherence Limit} \label{sect:coh}

A common technique used to estimate the \emph{minimum} error of a gate is to construct an amplitude and dephasing error channel for the duration of the gate and then calculate the average gate error. This can be done by representing those error channels in the Pauli superoperator form (Pauli Transfer Matrix, PTM, $R$) and then applying the formula described in Ref.~\cite{gambetta:2012}
\begin{equation}
    \epsilon = \frac{d}{d+1}\left(1-\frac{{\rm Tr}[R]}{d^2}\right)
\end{equation}
where $d=2^n$. Applying this formula for the 1 qubit gate we obtain
\begin{eqnarray}
    \epsilon_{1Q} = \frac{1}{6}\left(3-2e^{-t_g/T_2}-e^{-t_g/T_1}\right),
\end{eqnarray}
where $t_g$ is the length of the gate and $T_1$ and $T_2$ are the amplitude damping decay time and Ramsey decay time, respectively. For the two-qubit gate we take the tensor product of the superoperators for the qubits since the error channels are independent and obtain,
\begin{align}
\epsilon_{2Q} & =  \frac{1}{20}\Bigg(15-\sum_{i=0,1}\left[2e^{-t_g/T_{2,Qi}}+e^{-t_g/T_{1,Qi}}\right] \nonumber\\
&- e^{-t_g\left(1/T_{1,Q0}+1/T_{1,Q1}\right)} -4e^{-t_g\left(1/T_{2,Q0}+1/T_{2,Q1}\right)}\nonumber \\
&-2e^{-t_g\left(1/T_{1,Q0}+1/T_{2,Q1}\right)}-2e^{-t_g\left(1/T_{2,Q0}+1/T_{1,Q1}\right)}\Bigg). \label{eqn:coh2q}
\end{align}

\section{Purity RB}\label{sect:purity}
\begin{figure*}[!bpth]
\centering
\includegraphics[width=1.0\textwidth]{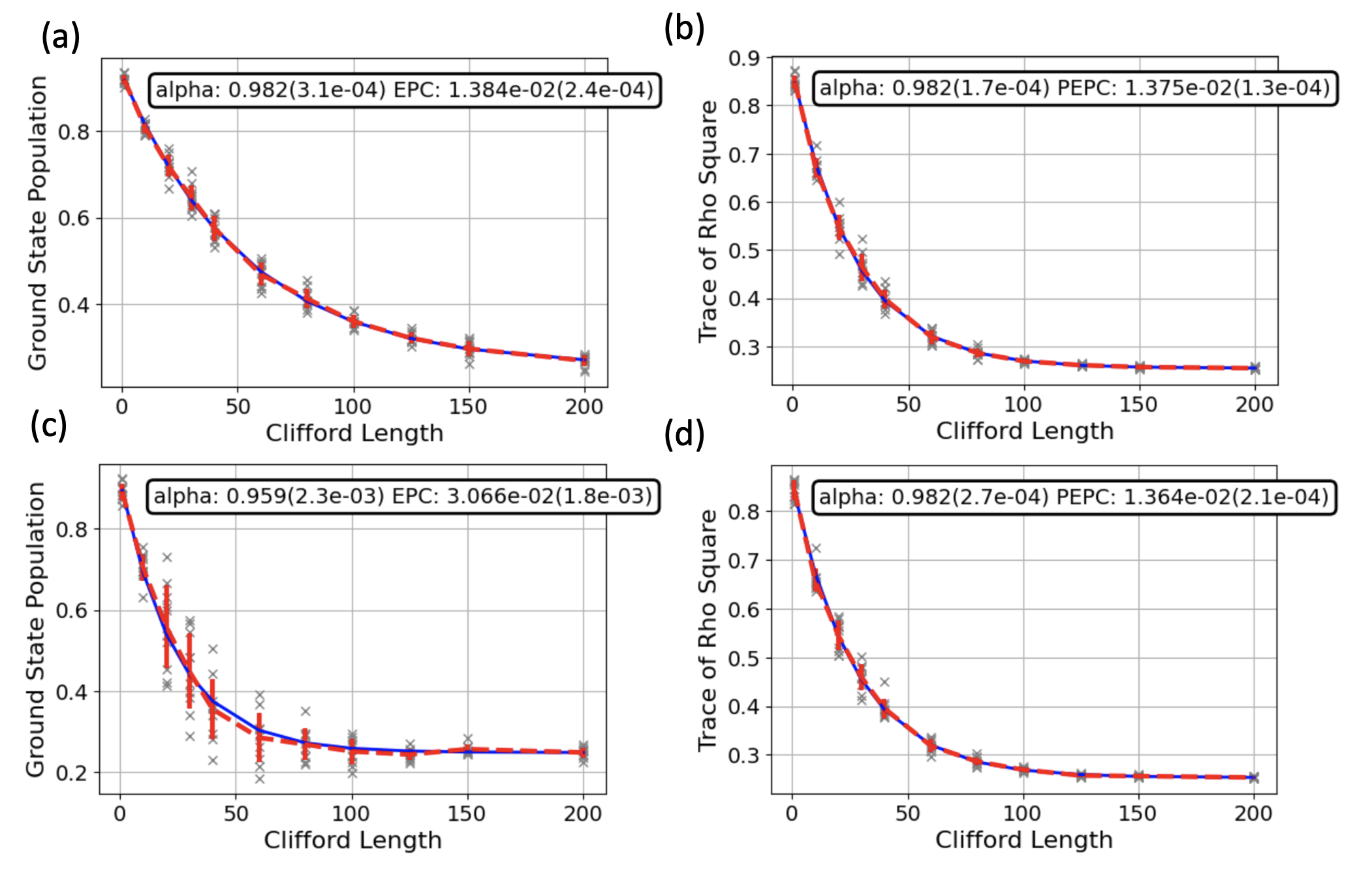}
\caption{Simulation of purity RB (b),(d) versus standard RB (a),(c) done in Qiskit~\cite{qiskit}. Here we choose $T_1$=$T_2$=40$\mu$s and for just the CX gate to be finite width of 300ns, there is also a 3\% measurement error. This has a coherence limit error of 1.35$\times10^{-2}$ per Clifford. For (a) and (b) there is only the $T_1$ and $T_2$ decohering errors. For (c) and (d) we add a coherent $X$ rotation error after each CX gate which degrades standard RB but does not change purity. Red lines are the average and standard deviation of the RB sequences, and blue lines are fits to the average.}
\label{FIG:puritysim}
\end{figure*}

The version of purity RB we use here is discussed in Ref.~\cite{mckay2016} and implemented in Qiskit-Ignis~\cite{qiskit}. For each Clifford sequences we append $3^n$ post-rotations to measure all $4^n$ Pauli expectation values required to calculate the purity of the state ${\rm Tr}[\rho^2]$. This can be seen from
\begin{eqnarray}
Tr[\rho^2] & = & {\rm Tr}[\sum_{i,j} a_i a_j P_i P_j] \nonumber \\
& = & \sum_{i} d a_i^2, 
\end{eqnarray}
since ${\rm Tr}[P_i P_j]$ is 0 if $i\ne j$ and $d$ otherwise (where $P_i$ is a Pauli operator). Also note,
\begin{eqnarray}
\langle P_i \rangle & = & {\rm Tr} [\rho P_i] \nonumber \\
& = & \sum_{i} d a_i, 
\end{eqnarray}
so
\begin{eqnarray}
{\rm Tr}[\rho^2] & = & \sum_i \frac{\langle P_i \rangle^2}{d}
\end{eqnarray}
If we assume depolarizing error then fitting ${\rm Tr}[\rho^2]$ vs Clifford length ($n$) to $A\gamma^{2n} + B$ we get that the incoherent error per gate is,
\begin{equation}
    \epsilon = \frac{3}{4} (1-\lambda^{1/n_2}),
\end{equation}
where $n_2$ is the number of 2Q gates per Clifford. We show some simulations of purity RB in FIG.~\ref{FIG:puritysim} illustrating how coherent errors degrade standard RB, but do not affect purity RB. 

\section{Standard techniques to characterize coherent errors}\label{HEAT}
\begin{figure}[thb!]    
\centering
\includegraphics[width=1.05\columnwidth]{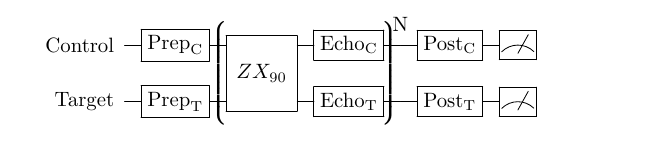}
\caption{The $k$th heat sequence ${\cal H}^N_k$}
\label{HEATseq}
\end{figure}

In \cite{HEATprx2020}, sequences were given to amplify $ZX$, $ZY$, $ZZ$, $IY$, and $IZ$ errors to the cross-resonance gate, as these errors are directly addressed during calibration of the amplitudes and phases of the cross resonance and target rotary pulse.  These sequences assume only that the unitary being amplified is nearly a $ZX_{\pi/2}$-rotation, and all other terms in the amplified unitary's effective Hamiltonian are relatively small. We can extend the sequences given in \cite{HEATprx2020} to include all other two-qubit Pauli errors by choosing pre, post, and echoing rotations from TABLE~\ref{heattable}.  The magnitude of Pauli error $\epsilon_{ij}$ can be obtained from fitting $\left<Z\right>_{C/T}$ following the appropriate HEAT sequences ${\cal H}_k^N$ of the form in FIG.~\ref{HEATseq} to a line in $N$ of slope $\alpha$. We choose numbers of repetitions N such that $N=0$ mod 4.  So for example, $\left<{\cal H}_1^N\right>_{\rm C}\simeq\alpha_1(\epsilon_{xi}+\epsilon_{xx})N/2$, or equivalently, $\epsilon_{xi}\simeq(\left<{\cal H}_1^N\right>_{\rm C}+\left<{\cal H}_2^N\right>_{\rm C})/2\alpha_1 N$.  We assume that all $\epsilon_{ij}$ are small compared to $\pi/2$ so that the resultant HEAT sequences fit to a line for up to $N\simeq20$.  In practice this assumption is nearly always reasonable after the execution of other standard calibration routines.

\section{Gate Set Tomography for a Stark $Z_{\pi/2}$}\label{GST}

\begin{figure*}[!tbh]
\centering
\includegraphics[width=0.75\textwidth]{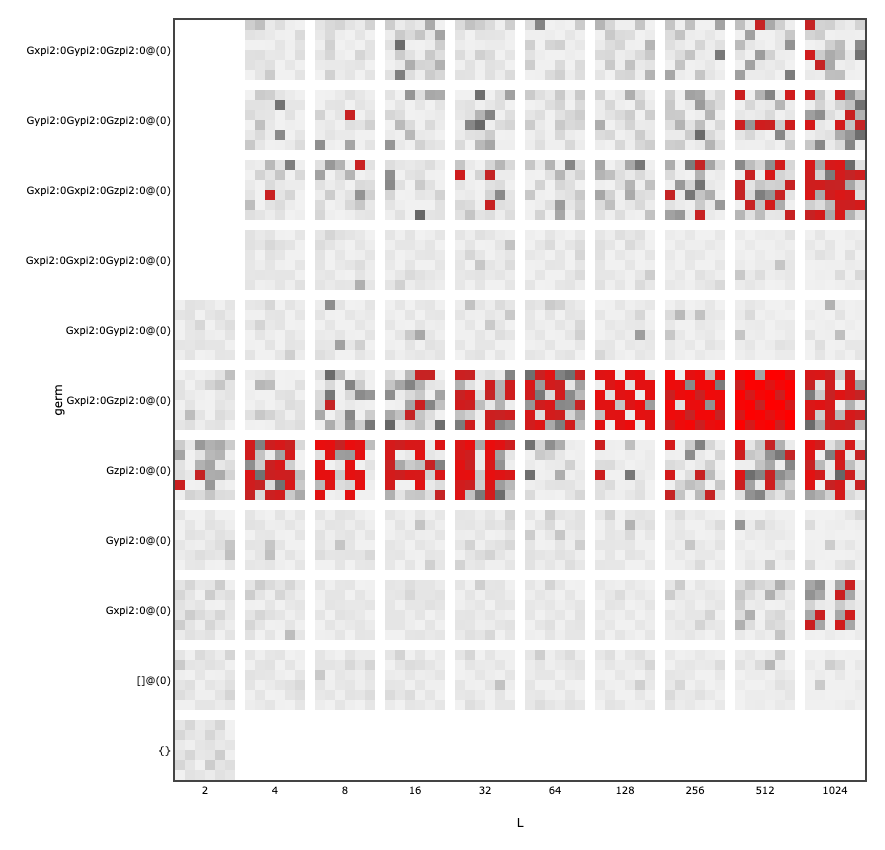}
\caption{Model violation in GST as reported by the loglikelihood ratio of the fit (red denotes experiments that are poorly described by the best fit Markovian model).  Each row describes a different germ sequence, a small sequence $X_{\pi/2}$,$Y_{\pi/2}$, and $Z_{\pi/2}$ gates.  These are then raised the power $L$. columns.  The 36 smaller squares inside each larger block denote different choices of the prep and measure operations.   }
\label{FIG:GST}
\end{figure*}

In this appendix we use the pyGSTi implementation of long-sequence GST \cite{Nielsen2021,pygsti} to show model violation for a simulation of the Stark $Z_{\pi/2}$ gate described in \S~\ref{Sec:Stark}.  Starting from Eq.~(\ref{Eq:Stark}), we assume a detuning of $\Delta/2\pi = -50$MHz, a phase $\phi = 0$,  and apply a square pulse of amplitude $\Omega$ and duration $t_g = 96$ns.  We then optimize $\Omega$ to perform a $Z_{\pi/2}$ gate, which results in $\Omega/2\pi \approx 16.25$MHz.  Integrating this Hamiltonian from $t = 0 \ldots t_g$ yields the unitary operator, in the resonant frame, of 
\begin{equation}
 U \approx 0.717 I -i(0.037 X + 0.027 Y + 0.695 Z),
\end{equation}
which has an average gate error of $1.5\times 10^{-3}$.  We assume the $X_{\pi/2}$ and $Y_{\pi/2}$ gates to be perfect gates with duration $35.55$ns.

We now simulate a GST experiment using this gate set, being careful to track the phase in the resonant frame, and obtain the results in FIG.~\ref{FIG:GST}.  GST uses a loglikelihood ratio test to determine to determine how plausible the raw data is based on the optimally fitted model.  Here the only model constraints are that each gate is described by a physical quantum operation, and that the simulation is Markovian.  For this experiment, GST has an extremely hard time finding a Markovian model to fit this data, even for very small numbers of pulses.    

\begin{table*}
\begin{tabular}{clcccccccclcccc}

\cline{1-1} \cline{3-4} \cline{6-7} \cline{9-10} \cline{12-13} \cline{15-15}
\multicolumn{1}{|c|}{Sequence} & \multicolumn{1}{l|}{} & \multicolumn{1}{c|}{${\rm Prep_{C}}$} & \multicolumn{1}{c|}{${\rm Prep_{T}}$} & \multicolumn{1}{c|}{} & ${\rm Echo_{C}}$ & \multicolumn{1}{c|}{${\rm Echo_{T}}$} & \multicolumn{1}{c|}{} & ${\rm Post_{C}}$ & \multicolumn{1}{c|}{${\rm Post_{T}}$} & \multicolumn{1}{l|}{} & Measure & \multicolumn{1}{c|}{$\alpha$} & \multicolumn{1}{c|}{} & \multicolumn{1}{c|}{Paulis} \\ \cline{1-1} \cline{3-4} \cline{6-7} \cline{9-10} \cline{12-13} \cline{15-15} 
\multicolumn{1}{|c|}{1} &  & $\mathbb{I}$ & $Y_{90}$ &  & $X$ & $\mathbb{I}$ &  & $X_{90}$ & $Y_{-90}$ &  & C & $\frac{\pi^2}{16}$ &  & $\frac{1}{2}(XI+XX)$ \\ \cline{1-1}
\multicolumn{1}{|c|}{2} &  & $\mathbb{I}$ & $Y_{-90}$ &  & $X$ & $\mathbb{I}$ &  & $X_{90}$ & $Y_{90}$ &  & C & $\frac{\pi^2}{16}$ &  & $\frac{1}{2}(XX-XI)$ \\ \cline{1-1}
\multicolumn{1}{|c|}{3} &  & $\mathbb{I}$ & $Y_{90}$ &  & $\mathbb{I}$ & $Z$ &  & $\mathbb{I}$ & $\mathbb{I}$ &  & T & $\frac{\pi^2}{16}$ &  & $\frac{1}{2}(ZZ+IZ)$ \\ \cline{1-1}
\multicolumn{1}{|c|}{4} &  & ${X}$ & $Y_{-90}$ &  & $\mathbb{I}$ & $Z$ &  & $X$ & $X_{-90}$ &  & T & $\frac{\pi^2}{16}$ &  & $\frac{1}{2}(ZZ-IZ)$ \\ \cline{1-1}
\multicolumn{1}{|c|}{5} &  & $\mathbb{I}$ & $Y_{90}$ &  & $\mathbb{I}$ & $Y$ &  & $\mathbb{I}$ & $\mathbb{I}$ &  & T & $\frac{\pi^2}{16}$ &  & $\frac{1}{2}(ZY+IY)$ \\ \cline{1-1}
\multicolumn{1}{|c|}{6} &  & ${X}$ & $Y_{90}$ &  & $\mathbb{I}$ & $Y$ &  & $X$ & $X_{90}$ &  & T & $\frac{\pi^2}{16}$ &  & $\frac{1}{2}(ZY-IY)$ \\ \cline{1-1}
\multicolumn{1}{|c|}{7} &  & $\mathbb{I}$ & $X_{90}$ &  & $\mathbb{I}$ & $X$ &  & $\mathbb{I}$ & $\mathbb{I}$ &  & T & 1 &  & $\frac{1}{2}(ZX+IX)$ \\ \cline{1-1}
\multicolumn{1}{|c|}{8} &  & ${X}$ & $X_{90}$ &  & $\mathbb{I}$ & $X$ &  & $X$ & $\mathbb{I}$ &  & T & 1 &  & $\frac{1}{2}(ZX-IX)$ \\ \cline{1-1}
\multicolumn{1}{|c|}{9} &  & $\mathbb{I}$ & $Y_{90}$ &  & $Y$ & $\mathbb{I}$ &  & $X_{-90}$ & $Y_{-90}$ &  & C & $\frac{\pi^2}{16}$ &  & $\frac{1}{2}(YI+YX)$ \\ \cline{1-1}
\multicolumn{1}{|c|}{10} &  & $\mathbb{I}$ & $Y_{-90}$ &  & $Y$ & $\mathbb{I}$ &  & $X_{90}$ & $Y_{90}$ &  & C & $\frac{\pi^2}{16}$ &  & $\frac{1}{2}(YI-YX)$ \\ \cline{1-1}
\multicolumn{1}{|c|}{11} &  & $\mathbb{I}$ & $Y_{90}$ &  & $Y$ & $Y$ &  & $Y_{-90}$ & $X_{-90}$ &  & C & $\frac{1}{2}$ &  & $\frac{1}{2}(YY-XZ)$ \\ \cline{1-1}
\multicolumn{1}{|c|}{12} &  & $\mathbb{I}$ & $Y_{-90}$ &  & $X$ & $Z$ &  & $Y_{90}$ & $X_{-90}$ &  & T & $\frac{1}{2}$ &  & $\frac{1}{2}(YY+XZ)$ \\ \cline{1-1}
\multicolumn{1}{|c|}{13} &  & $\mathbb{I}$ & $Y_{90}$ &  & $X$ & $Y$ &  & $X_{90}$ & $X_{-90}$ &  & C & $\frac{1}{2}$ &  & $\frac{1}{2}(XY-XZ)$ \\ \cline{1-1}
\multicolumn{1}{|c|}{14} &  & $\mathbb{I}$ & $Y_{-90}$ &  & $Y$ & $Z$ &  & $X_{-90}$ & $X_{-90}$ &  & T & $\frac{1}{2}$ &  & $\frac{1}{2}(XY+XZ)$ \\ \cline{1-1}
\multicolumn{1}{|c|}{15} &  & $Y_{90}$ & $\mathbb{I}$ &  & $Z$ & $\mathbb{I}$ &  & $X_{90}$ & $\mathbb{I}$ &  & C & 1 &  & $ZI$ \\ \cline{1-1}
\multicolumn{1}{l}{} &  & \multicolumn{1}{l}{} & \multicolumn{1}{l}{} & \multicolumn{1}{l}{} & \multicolumn{1}{l}{} & \multicolumn{1}{l}{} & \multicolumn{1}{l}{} & \multicolumn{1}{l}{} & \multicolumn{1}{l}{} &  & \multicolumn{1}{l}{} & \multicolumn{1}{l}{} & \multicolumn{1}{l}{} & \multicolumn{1}{l}{}
\end{tabular}
\caption{Summary table of generalized HEAT sequences to amplify coherent errors.}
\label{heattable}
\end{table*}

\clearpage
\bibliography{references}
\end{document}